%% file: manuscript_PLOS.tex
\RequirePackage{snapshot}
\documentclass[10pt,letterpaper]{article}
\usepackage[top=0.85in,left=2.75in,footskip=0.75in]{geometry}

\usepackage{amsmath,amssymb}
\usepackage{mathrsfs}

\usepackage{xr-hyper}
\usepackage{changepage}

\usepackage{textcomp,marvosym}

\usepackage{cite}

\usepackage{nameref,hyperref}

\usepackage[right]{lineno}

\usepackage[nopatch=eqnum]{microtype}
\DisableLigatures[f]{encoding = *, family = * }

\usepackage[table]{xcolor}

\usepackage{array}

\newcolumntype{+}{!{\vrule width 2pt}}

\newlength\savedwidth

\raggedright
\usepackage[aboveskip=1pt,labelfont=bf,labelsep=period,justification=raggedright,singlelinecheck=off]{caption}
\renewcommand{\figurename}{Fig}

\makeatletter
\renewcommand{\@biblabel}[1]{\quad#1.}
\makeatother

\usepackage{lastpage,fancyhdr,graphicx}
\usepackage{epstopdf}
\renewcommand{\headrulewidth}{0pt}
\renewcommand{\footrule}{\hrule height 2pt \vspace{2mm}}
\fancyheadoffset[L]{2.25in}
\fancyfootoffset[L]{2.25in}
\newcommand{\expect}[1]{\mathrm{E}\left[#1\right]}
\newcommand{\refeqn}[1]{Eq. \ref{#1}}
\newcommand{\refapdx}[1]{Appendix \ref{#1}}

\DeclareMathOperator*{\argmax}{arg\,max}
\DeclareMathOperator*{\argmin}{arg\,min}

 \DeclareMathOperator{\Var}{Var}

\newcommand{\ts}{\textsuperscript}

\begin{document}

\vspace*{0.2in}
\begin{flushleft}
{\Large
\textbf{Modulation Analysis with Higher-Order Spectra} 
}
\newline

Christopher K. Kovach\textsuperscript{1*},
Sukhbinder Kumar\textsuperscript{2},
\\
\bigskip
\textbf{1} Department of Neurosurgery, University of Nebraska Medical Center, Omaha, NE, USA
\\
\textbf{2} Department of Neurosurgery, University of Iowa Hospitals and Clinics, Iowa City, IA, USA
\\
\bigskip

* ckovach@unmc.edu

\end{flushleft}


\begin{abstract}

\input{abstract.tex}

\end{abstract}


\section*{\label{sec:level1}Introduction}

Higher order spectra (HOS) are frequency-domain representations of time-shift invariant higher-order cumulants \cite{brillinger1965introduction,nikias1993signal}. 
Along with time invariance, important properties of HOS include that they are additive for independent additive processes, vanish for stationary Gaussian processes, and retain information about phase.
In practical terms, these properties mean that, given enough data, HOS can be estimated without bias in Gaussian noise under complete uncertainty of timing.
Information in HOS may be used to identify non-Gaussian features and non-minimum-phase systems under high levels of additive Gaussian noise.
A more detailed overview of relevant properties is given in \refapdx{APDX:HOS}.

Although HOS provide a mathematically elegant and comprehensive description of the cross-frequency dependencies that arise from nonlinear and non-Gaussian processes, adoption of HOS-related techniques has been hampered by a perceived obscurity of real-world meaning \cite{ brillinger1965introduction,akaike1966note}.
Nevertheless, the literature describing various practical applications of HOS has grown steadily, if slowly, with notable applications to biomedical signal processing \cite{chua2010application}. 
Progress on questions of interpretation can be found in recent work exploring the relationships between the third-order spectrum (bispectrum) and measures of phase-amplitude coupling widely adopted in electrophysiological literature  \cite{Sheremet4218, Kovach2018,avarvand2018localizing, sheremet2020investigation, zandvoort21defining,kovach2026interpreting}, and in new strategies and techniques for exploiting the information in HOS to identify and recover transient or harmonically related signal components \cite{Kovach2019,giannakis1992unifying,bendory2018}. 
The present work seeks a similar development for the blind identification of characteristic patterns of modulation from the  fourth-order spectrum, also known as the \emph{trispectrum}. 

\subsection*{\label{sec:Prob}Statement and justification of the problem}

For simplicity, it will be assumed that all signals are univariate and real valued, though the extension to multivariate and complex signals is straightforward \cite{Kovach2019}. 
The problem can be formalized as a nonstationary system driven by a stationary white noise orthogonal increment process, $Z$: 
\begin{align}
\label{EQ:WoldCramer}
\begin{split}
	x(t) = \int H(\omega,t-t_0)e^{i\omega t}\mathop{dZ(\omega)}
\end{split}
\end{align}
Broadly stated, the goal is to identify the statistical properties of nonstationarity in $H$, where $H$ is subject to the following assumptions.

\subsubsection*{Conditional nonstationarity}
It will be assumed that the system, $H$, is nonstationary, conditioned on a completely unknown random time delay, $t_0$.
This describes the common scenario in which a nonstationary signal is measured within an arbitrarily fixed timeframe, whose origin has no anticipated relationship to the nonstationarity of the signal \cite{antoni2006spectral}.
In the frequency domain, the $K^\mathrm{th}$-order moment takes the form
\begin{align}
\label{EQ:Mdef}
M_0^K = \mathrm{E}\left[X(\omega_1)X(\omega_2)\dots X(\omega_K)\right]
\end{align}
which under a random time shift, $t_0$, becomes
\begin{align}
\label{EQ:Mts}
M^K = M_0^K\mathrm{E}\left[e^{-i\left(\sum_{k=1}^K\omega_k\right)t_0}\right] 
\end{align}
If one supposes $t_0$ to be drawn from a distribution with maximum entropy then
\begin{align}
\label{EQ:Mtvanish}
\begin{array}{lll}
M^K =  0&\text{ if }& \sum_{k=1}^K{\omega_k} \ne 0
\end{array}
\end{align}
Time-varying moments vanish as a consequence of the uniform distribution of phase induced in \refeqn{EQ:Mts} by random time shifts; moments which do not vanish must therefore be time-shift invariant.
The second-order shift-invariant moment is the power spectrum;
\emph{higher-order spectrum} refers to a shift-invariant moment (or cumulant) of order greater than 2, represented in the frequency domain.
For both stationary and conditionally non-stationary processes, HOS are non-vanishing higher moments.

\subsubsection*{Boundedness}
Under uncertainty in $t_0$ the distribution of $x$ effectively becomes equal to the marginal distribution over time, provided such a distribution exists.
If the marginal distribution exists, a conditionally nonstationary system, in fact, meets the definition of a stationary system.
It will be assumed here that the trispectrum of the process exists and is finite, and therefore the process is fourth-order stationary.

\subsubsection*{Consistency}
A system that randomly emits a single isolated, finite, non-recurring transient waveform in a background of additive Gaussian noise formally counts as both bounded and conditionally nonstationary; but statistics of the waveform cannot be measured, except within a window of finite duration that includes the waveform, violating the spirit of time-invariance. 
Thus the assumption of boundedness is complemented with a further notion of stability over repeated observation, which guarantees the availability of consistent estimators for relevant non-trivial statistics.
The statistical term will be misappropriated here by referring to this property of the system as consistency. 

\subsubsection*{Ergodicity}
The assumptions of boundedness and consistency are closely related to the standard assumption of ergodicity: given enough time, an ergodic system eventually emits every finite sequence within the ensemble. 
Ergodicity justifies the assumption that the behavior of a system observed over a sufficiently long time gives a statistically useful sampling of the totality of its possible states. In other words, the behavior of the system exhibits consistency in the manner just described.

\subsection*{Aside on stationarity}
The concept of conditional non-stationarity as non-stationarity subjected to a random time shift is useful for clarifying what is meant by stationarity. 
Three key points are: 1) there is nothing fundamentally improper about measuring the time-invariant moments of nonstationary systems, provided the relevant moments are bounded. 
2) For ergodic conditionally nonstationary systems, time-invariant moments give a complete description of the bounded statistics of the underlying ensemble.
3) Stationarity does not imply the absence any evolution over time whatsoever. 
In fact, HOS may be interpreted as describing the dynamics of lower order statistics. 
Thus, it will be possible to recover the spectrum of conditional second-order nonstationarities from time-invariant fourth-order statistics.
This point becomes a useful way to bootstrap some insight into the meaning of higher-order spectra, as demonstrated next.



\section*{A Statistical View of Amplitude Modulation }

Given the model in \refeqn{EQ:WoldCramer}, Amplitude modulation (AM) will be identified with a conditionally nonstationary system for which $H$ factors into a stationary term, the ``carrier'', $g$, and a  time-varying term, the ``envelope'', $y$ such that 
\begin{align}
\label{EQ:AM_model}
H(\omega,t) = y(t)G(\omega)
\end{align} 
The envelope, $y$, is assumed independent of the carrier, and for the purpose of characterizing its properties, we may model it as the result of a separate stochastic process, leading to a doubly stochastic signal model:
\begin{align}
y(t) = \int{F(\omega)e^{i\omega t}\mathop{d\Gamma(\omega)}}
\end{align}
with $\Gamma$ an orthogonal increment white noise process.
We may further consider mixtures of independent processes of the type 
\begin{align}
y(t) = \sum_k \int{F_k(\omega)e^{i\omega t}\mathop{d\Gamma_k(\omega)}}
\end{align}
For simplicity, $Z$ will be assumed a Wiener process (i.e Gaussian white noise increments), while increments of $\Gamma_i$'s need not be Gaussian and might, for example, be a Poisson process. 
In developing the modulogram, it will also be assumed that $y$ has a non-vanishing mean, as would be the case for non-negative $y$, so that the carrier band is preserved in $x$. 

The problem of identifying amplitude modulation is that of distinguishing $H_0 =  G(\omega)$ from $H_1 = y(t)G(\omega)$, with non-constant y, and otherwise identifying $f_i$'s to the extent possible. 
Modulation by a real-valued envelope induces correlations between sidebands surrounding the carrier frequency.
As explained next, the trispectrum can be taken as a measure linear dependence between power at different frequencies, suitable for detecting the relevant cross-frequency correlation.

\begin{figure} 
\includegraphics[width=3in]{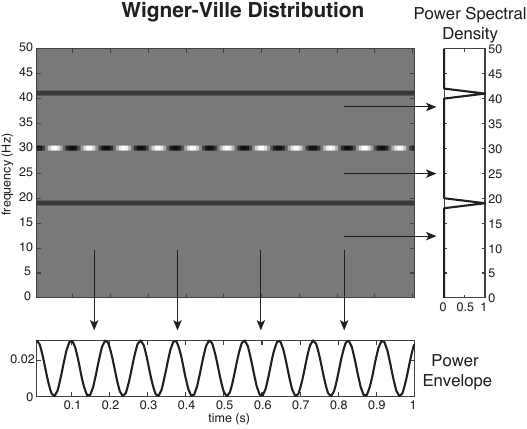}
\caption{\label{fig:wvd_examp} 
The meaning of interference in the Wigner-Ville pseudo distribution. (WVD).
The WVD gives a comprehensive representation of the quadratic properties of a signal in time and frequency, but the phenomenon of  ``interference'' is considered to detract from its usefulness as a time-frequency distribution.
Here a signal composed of two pure sinustoids, at 19 Hz and at 41 Hz, combined,  exhibit interference in the WVD at the midpoint of 30 Hz. 
Energy at 30 Hz does not manifest in the power spectral density of the signal, which integrates the WVD over time.
Yet the signal admits an equally valid interpretation as a pure 30 Hz tone modulated by a 5.5 Hz sinusoidal envelope (resulting in the 11 Hz power envelope shown here on integrating over frequency). 
The WVD encompasses both interpretations, and the identification of ``interference'' at 30 Hz is less an artifact of the WVD than a reflection of ambiguity in the signal.
}
\end{figure}

\subsection*{\label{sec:WVD}Trispectrum as a measure of linear cross-frequency dependencies of power}
Higher-order spectra may be regarded as representing dynamic properties of lower-order statistics, which becomes useful as a handle for developing some intuition about them \cite{tick1961estimation,akaike1966note,kovach2026interpreting}.
In particular,  linear coupling of power across frequencies is naturally expressed as a fourth-order spectrum. 
While such coupling might be estimated by extracting power from a standard time-frequency decomposition or filter bank, these approaches are sensitive to the choice of frequency resolution in the decomposition or filter bandwidth and biased in the presence of Gaussian noise.
Both problems are eliminated when adopting the Wigner-Ville pseudo-distribution (WVD) as the TFD of choice.
 The WVD is defined as 
\begin{align}
\label{EQ:WVDdef}
\begin{split}
\mathcal{W}&\left\{x\right\}(\omega,t) 
=\int{x\left(t+\frac{\tau}{2}\right)x^*\left(t-\frac{\tau}{2}\right)e^{-i\omega\tau}\mathop{d\tau}} 
\end{split}
\end{align}
Most other commonly used quadratic time-frequency distribution, such as the spectrogram, scalogram and bandpass filtering with rectification, can be obtained by applying a smoothing kernel to the WVD \cite{Cohen1989,hlawatsch1992linear}. 
Because the WVD has the least smoothing bias among alternative TFDs \cite{aviyente2004minimum}, it forms a natural starting point to study the temporal dynamics of power in a signal. 

Treating the WVD as a multivariate time series, with frequency bands taking the place of channels, we observe the linear dependence  of power across frequencies. 
Because the WVD is a quadratic time-frequency distribution, its time-invariant moments of order $k$ correspond to time-invariant moments of order $2k$ for the original process.
This point is easily grasped for $k=2$ by considering the Fourier transform of the WVD along the time dimension:
\begin{align}
\label{EQ:MSDdef}
\begin{split}
&\mathcal{M}\left\{x\right\}(\omega,\eta)  =\\
&\int{\int{x\left(t+\frac{\tau}{2}\right)x^*\left(t-\frac{\tau}{2}\right)e^{-i\omega\tau-i\eta t}\mathop{d\tau}}
\mathop{dt}} \\
&=X\left(\omega+\frac{\eta}{2}\right)X^*\left(\omega-\frac{\eta}{2}\right)
\end{split}
\end{align}
This spectrum clearly must contain information about the modulation of power at frequency, $\omega$, with a modulating frequency, $\eta$, and following  \cite{atlas2003joint} will be referred to as the \emph{modulation spectrum}.
Although the magnitude of the modulation spectrum has been proposed as a tool to identify modulation \cite{atlas2003joint}, it does not appropriately quantify the \emph{correlation} between frequencies induced by modulation and therefore cannot, by itself, give us exactly what we want.
For this, we need the full cross spectrum of the WVD:
\begin{align}
\label{EQ:MSmom}
\begin{split}
\mathrm{M^2} &\left\{\mathcal{W}\{x\}\right\}(\omega_1,\omega_2,\eta)  \\
=&\mathrm{E}\left[\mathcal{M}(\omega_1,\eta)\mathcal{M}^*(\omega_2,\eta)\right]\\
=&\mathrm{E}\left[X\left(\omega_1+\frac{\eta}{2}\right)X^*\left(\omega_1-\frac{\eta}{2}\right)X^*\left(\omega_2+\frac{\eta}{2}\right)X\left(\omega_2-\frac{\eta}{2}\right)\right]
\end{split}
\end{align}
But this is simply a reparameterization of the fourth-order spectrum of $x$, $M^4\{x\}$.
The trispectrum therefore quantifies linear relationships across frequency bands in the WVD. 
Retaining the WVD as an intermediate step adds unnecessary computational expense, and it makes more sense to compute the trispectrum directly.
Nevertheless, its derivation from the WVD highlights a simple and intuitive interpretation for the trispectrum.


As reviewed in \refapdx{APDX:HOS}, the trispectrum, like other HOS, contains enough information to completely recover the phase spectrum of any transient and deterministic $x$, up to a time shift and a constant term in the phase (for real-valued signals, meaning sign), implying that the WVD must likewise be an invertible transform. 
This contrasts with time-localized power computed via the squared modulus of the STFT or squared output of a filter bank, which,  as a consequence of discarding phase, is not invertible. 
Information about phase in the WVD takes the form of ``interference,'' which is lost to the smoothing implicit in conventional methods for recovering time-varying power through windowed Fourier TFDs or  bandpass filtering.
An illustration of the meaning of interference is given in {\bf Figure \ref{fig:wvd_examp}} for a signal composed of two pure tone sinusoids. 
One might, in this example, regard interference as a non-physical distortion of the true signal model; but, in fact, it appropriately conveys the equivalence between two signal models that cannot be distinguished with the available information: as a sum of two interfering sinusoids or a single sinusoid modulated by a real-valued sinusoidal envelope.

\begin{figure} 
\includegraphics[width=3in]{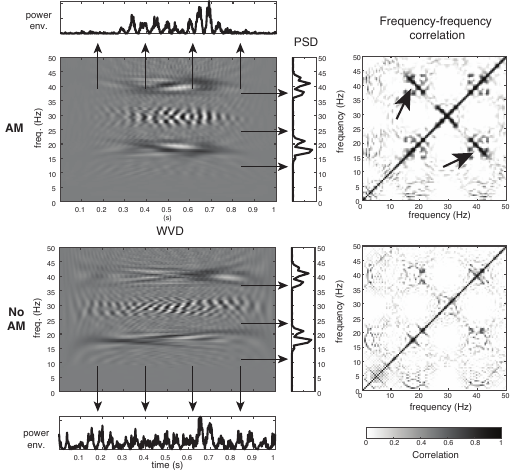}
\caption{\label{fig:wvd_examp_random} 
Cross-frequency correlation in the WVD induced by amplitude modulation (AM). Two example signals were generated; the first (\emph{top row}) was created by applying a real-valued 8-14 Hz filtered Gaussian noise envelope to a 30 Hz tone. The second (\emph{bottom row}) was generated with the same power spectrum as the first by randomizing Fourier domain phase while retaining the same amplitude spectrum. 
In the first case, AM is revealed by symmetry in bands on either side of the carrier frequency within the WVD (\emph{upper left panel}), as shown by a negatively sloped diagonal band in the correlation between bands (\emph{arrows, upper right panel}).
In the second case, upper and lower bands are independent due to phase randomization, and a similar pattern of correlation is not observed (\emph{lower right panel}).
}
\end{figure}

\subsection*{Identifying Amplitude Modulation}

The ambiguity in the foregoing example can be resolved as a question about the statistical relationship between side bands. 
Because the spectrum of an amplitude modulated signal is the convolution between the carrier spectrum and the modulating signal spectrum, AM involving a real-valued envelope induces a positive correlation between frequencies offset by equal amounts on either side of the carrier band 
({\bf Fig. \ref{fig:wvd_examp_random}}). 
A similar phenomenon exists for frequency modulation (FM); 
but in the FM case, the correlations are negative \cite{luo2006concurrent,Kovach2018}.
This is so because the total power of an FM signal remains nearly constant within its bandwidth,  but power meanders through sub-bands according to the modulation of instantaneous frequency. For total power to remain approximately constant, when power increases in one sub-band it must tend to decrease in others, hence a negative correlation between sub-bands.
AM and FM may be distinguished by the direction of the statistical relationship between neighboring bands in the WVD, with the relevant information contained in the phase  of the trispectrum.

%

For the doubly stochastic model in \refeqn{EQ:AM_model}, conditioned on $t_0$ the Wigner-Ville distribution is
\begin{align}
\label{EQ:WVD_doubly_stochastic}
\begin{split}
&\expect{\mathcal{W}\{x\}(\omega,t) } =\\
\sigma^2_z&\int{\expect{\mathcal{W}\{y\}\left(\eta-\omega, t - t_0\right)}\left|G(\eta)\right|^2 d\eta}
\end{split}
\end{align}
The WVD of $y$ is therefore preserved in frequency shifted form in the limit of a pure tone carrier, $|G(\omega)|^2 \rightarrow \delta(\omega-\omega_0)$.
In general, the distortion becomes negligible when the bandwidth of the carrier is narrow compared to the characteristic time scale of autocorrelation in $y$, reflected in the smoothness of its spectrum. 
The trispectrum in \refeqn{EQ:MSmom} likewise contains a frequency shifted copy of the trispectrum of $y$ in this limit:
\begin{align}
\label{EQ:MSymom}
\begin{split}
\mathrm{M^2} &\left\{\mathcal{W}\{x\}\right\}(\omega_1,\omega_2,\eta) =\\
&\mathrm{M^2} \left\{\mathcal{W}\{y\}\right\}(\omega_1-\omega_0,\omega_2-\omega_0,\eta) 
\end{split}
\end{align}

\subsection*{The Trispectral Modulogram}
\label{SECTION:modulogram}
One might suppose that information of greatest relevance for identifying AM-induced correlations across frequency bands is contained in the plane, $\eta=0$ in \refeqn{EQ:MSmom}, which 
gives
\begin{align}
\label{EQ:trispect_degen}
\mathrm{M^2} &\left\{\mathcal{W}\{x\}\right\}(\omega_1,\omega_2,0) =\mathrm{E}\left[\left|X(\omega_1)\right|^2\left|X(\omega_2)\right|^2\right]  
\end{align}
Similarly, one might hope to recover the power spectrum of AM within a given band by setting $\omega_1 = \omega_2$ in \refeqn{EQ:MSmom} \cite{atlas2003joint}, which is the expectation of what is conveyed by the squared magnitude of the modulation spectrum in \refeqn{EQ:MSDdef}, which is also equivalent to \refeqn{EQ:trispect_degen} by the symmetry of the trispectrum. 
This strategy has been used in both forms elsewhere \cite{atlas2003joint,capdevielle1996blind},  but by restricting attention to degenerate subdomains of the trispectrum (degenerate here meaning that subsets of frequencies sum to zero; see the discussion in \refapdx{APNDX:mometcum}) it is limited in several respects: first, it introduces noise-dependent bias in the moment spectrum as a result of passing Gaussian noise through a symmetric nonlinearity. 
Although this bias might be corrected by adding the lower-order terms of the cumulant expansion to \refeqn{EQ:trispect_degen}, estimation error remains inflated compared to estimators within the non-degenerate domains of HOS. 
Finally, and most significantly, by limiting the estimate to the 0 Hz band of the WVD cross spectrum, it introduces a conceptual contradiction: it attempts to quantify variability over time by a coefficient that represents the constant part of the signal.  
In practice, this means that the resulting estimators convey correlations of power within a lowpass range of frequencies determined by the broadband bias of the estimator. 

To understand the inflation of estimation error, one may note that the variance of a product of 4 independent samples from a Gaussian distribution, $Z\sim N(0,\sigma^2)$,  is $\Var\left\{Z_1Z_2Z_3Z_4\right\}/\sigma^8 = 1$; for the squared product of two independent samples, it rises to $\Var\left\{Z_1^2Z_2^2\right\}/\sigma^8 = 8$,  while the fourth power of a single sample adds another order of magnitude: $\Var\left\{Z_1^4\right\}/\sigma^8 = 96$. 
For the same reason, the estimators within non-degenerate domains of HOS, which for a Gaussian time invariant process involves the product of 4 independent complex Gaussian variables, behave favorably with respect to bias and error  (with respective variances of $1\sigma^8$, $3\sigma^8$ and $20\sigma^8$ if $\mathrm{E}\left[\left|Z\right|^2\right] = \sigma^2$).


\subsubsection*{The Diagonal Slice}

\begin{figure*} 
\includegraphics[width=4in]{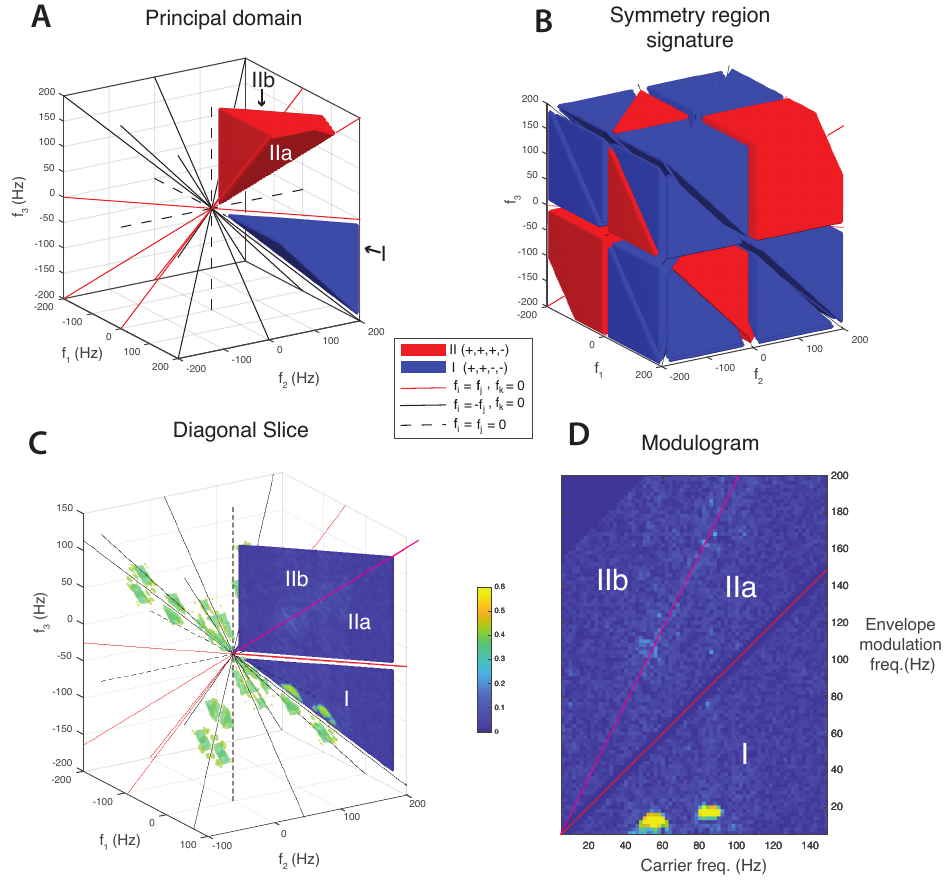}
\caption{\label{fig:diagonal_slice} 
Structure of the trispectrum. {\bf A,B}: Symmetry region signatures within the principal domain (\emph{A}) and throughout the trispectrum (\emph{B}). 
In the trispectrum, signs of all frequencies in nondegenerate regions either sum to 0 (\emph{blue}) or $\pm$ 2 (\emph{red}).
The trispectrum of a narrowband signal with fractional bandwidth less than 1 vanishes outside regions with balanced signature (\emph{blue}). Signatures in the legend are for the respective principal domains.
{\bf C,D}: Normalized trispectrum (tricoherence) for a signal containing two amplitude modulated components, at carrier frequencies of $50- 60 \text{ Hz}$ and  $80- 90 \text{ Hz}$ for the full trispectral domain, with the principal diagonal slice filled in \emph{C} and reparameterized as the modulogram in \emph{D}. Note that Region IIa depicted here is equivalent to region IIa depicted in panel \emph{A} by symmetry of the auto-trispectrum.} 
\end{figure*}

Amplitude modulation results in a spectrum with sidebands that are symmetrically offset from a carrier frequency.
To better describe this situation, we may re-parameterize  $\omega_1$ and $\omega_2$ in \refeqn{EQ:MSmom} according to the carrier frequency, $\omega$, and sideband distance, $\beta$, as $\omega_1 = \omega - \beta/2$ and $\omega_2 = \omega + \beta/2$, giving
\begin{align}
\label{EQ:MSmomreparam}
\begin{split}
\mathrm{M^2}&\left\{\mathcal{W}\{x\}\right\}(\omega,\beta,\eta)  = \\
=\mathrm{E}&\left[X\left(\omega+\frac{\eta-\beta}{2}\right)X^*\left(\omega -\frac{\eta+\beta}{2}\right)\right. \\
\times & \left.X^*\left(\omega+\frac{\eta+\beta}{2}\right)X\left(\omega-\frac{\eta-\beta}{2}\right)\right]
\end{split}
\end{align}
The carrier of an AM signal is typically understood to be narrowband, with a bandwidth that limits the distortion of $y$.
It is also often assumed that the envelope modulating signal falls in a low frequency baseband, with no overlap in the bands of the carrier and modulating signals.
The distinction between the modulating signal and carrier becomes natural under this condition, known as Bedrosian's condition, because the product of the former with the analytic envelope of the latter gives an analytic representation of the composite signal \cite{bedrosian1963product}. 
When Bedrosian's condition holds, the trispectrum vanishes outside the domain with balanced frequency signature (region I in {\bf Fig. \ref{fig:diagonal_slice}});
on the other hand, when it does not hold, the distinction between carrier and modulating signal becomes muddled, as the modulating signal contaminates the analytic phase of the product.
It is also often assumed that the modulating signal is non-negative, which allows the analytic envelope to be identified with the non-negative modulus of the analytic signal. 
Again, this requirement is not a strict one in the present discussion; however, it will be assumed that the modulating signal is lowpass so that energy is retained within the original band of the carrier.
Energy present around $0$ Hz in the modulating signal preserves a portion of the carrier within the original band, between upper and lower sidebands (illustrated in {\bf Fig. \ref{fig:band_illo}}).

The residual presence of the carrier motivates the use of a 2-dimensional subdomain of the trispectrum that conveys the three-way dependence between upper and lower sidebands and the carrier band, corresponding to the plane with $\beta-\eta=0$, which gives
 \begin{align}
\label{EQ:DSdef}
\begin{split}
\mathrm{M_{DS}^4}&(\omega,\eta)  = \mathrm{E}\left[X^2\left(\omega\right)X^*\left(\omega -\eta\right)X^*\left(\omega+\eta\right)\right]
\end{split}
\end{align}
The parameterization according to center frequency and sideband distance leads to the interpretation of the trispectrum as spectrally resolved power in the modulating signal for a carrier at the given center frequency, schematically illustrated in {\bf Figure \ref{fig:band_illo}}. 
The subdomains in the trispectrum for which two of the four frequency arguments are equal, or equal with inverted sign under complex conjugation, will be referred to as the \emph{diagonal slice}. 
%
%

The full trispectrum contains 6 such slices, due to symmetry, but for an autospectrum, one need only consider the single slice within the principal domain of the trispectrum (see {\bf Fig. \ref{fig:diagonal_slice}}).
The three-dimensional space of the full trispectrum grows as the cube of signal length, which may become computationally inconvenient, even when analysis windows are kept relatively short.
Narrowing attention to the diagonal slice has a computational advantage in addition to yielding a more tractable and easily interpreted representation of the relevant part of the trispectrum.
As long as energy is retained within the carrier band, it will be possible to identify AM within this subdomain of the trispectrum.

\section*{Additive Decomposition}

The modulogram represents a windowing of the 4\ts{th}-order spectrum, restricted to the diagonal slice. 
A recently described technique for decomposing HOS of any order (HOSD) \cite{Kovach2019}, is applicable to windowed HOS of this type.
The motivation and implementation of the algorithm is explained in \refapdx{APDX:HOSD_Algorithm}.

\begin{figure*} 
\includegraphics[width=\textwidth]{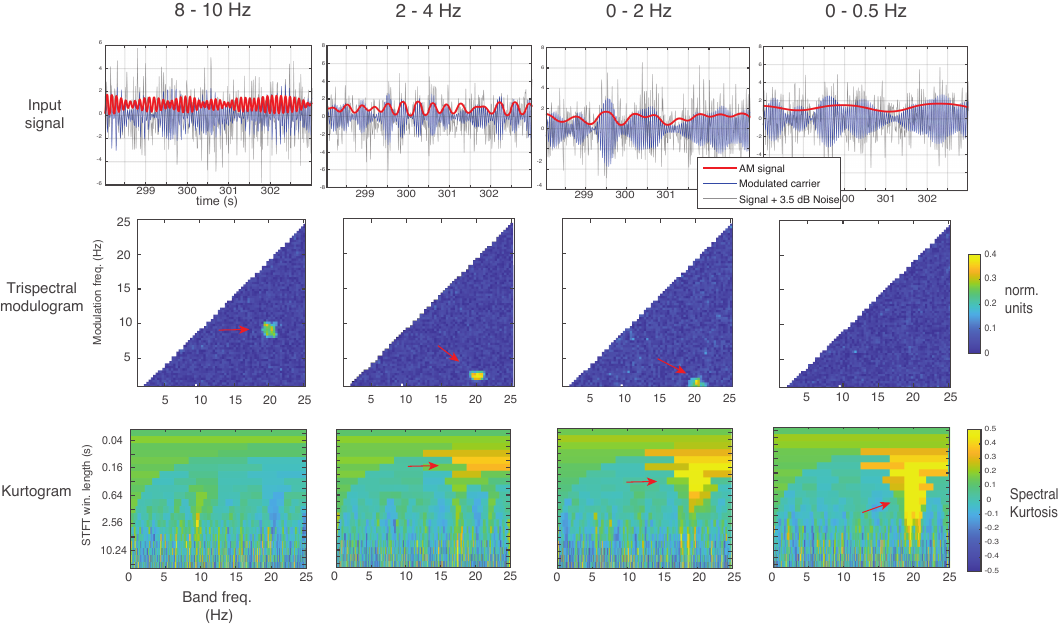}
\caption{\label{fig:mdg_ktg_comp} 
Comparison of the trispectral modulogram and the kurtogram in the detection of amplitude modulation. 
Simulated test signals were composed of a non-negative AM envelope (\emph{top row, red}) applied to a 20 Hz $\pm$1 Hz filtered Gaussian carrier  (\emph{blue}) to which 3.5 dB additive Gaussian white noise  (\emph{gray}) was added. 
The envelope was generated by passing filtered and standardized Gaussian noise through a  static sigmoidal nonlinearity, using 4 different filter band (\emph{columns}).
The modulogram (\emph{second row}), computed with with a 3 s analysis window, and the kurtogram (\emph{third row}) at 20 different STFT window scales are compared in their ability to identify amplitude modulation within the test signals.
The effect of amplitude modulation is evident in the modulograms for all test signals, except the one whose AM timescale falls below the highpass threshold (0.67 Hz) imposed on the trispectral estimator (\emph{rightmost column}). 
In contrast, SK exhibits a positive relationship between sensitivity and fractional bandwidth of the AM signal, because it lacks sensitivity to modulation at scales shorter than the STFT analysis window.
SK fails completely to identify narrow band amplitude modulation (with low fractional bandwidth) at any scale (\emph{leftmost column}) due to inadequate spectral resolution of the short analysis windows.  
Performance of SK improves as the timescale and fractional bandwidth of the AM signal progressively increases (\emph{left to right}).
}
\end{figure*}

\begin{figure} 
\includegraphics[width=3in]{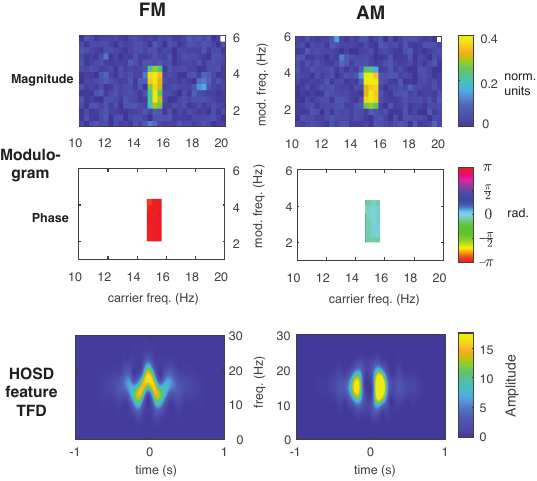}
\caption{\label{fig:am_vs_fm} 
Distinguishing frequency (FM) from amplitude (AM) modulation with modulogram phase. 
Modulograms were computed for two simulated signals, one frequency modulated (\emph{left column}) and the other amplitude modulated (\emph{right column}).
Magnitude modulograms (\emph{top row}) are nearly identical, but phase modulograms diverge by $\pi$ rad. (\emph{2\ts{nd} row}), with a phase of $\pi$ in the FM case and $0$ in the AM case, reflecting the fact that power in upper and lower sidebands is in-phase for an AM signal and anti-phase for FM. 
This fact is also reflected in the time-frequency plots of the first HOSD feature recovered from each signal (\emph{3\ts{rd} row}). 
In each case, the modulating signal was generated from 2-4 Hz filtered Gaussian noise with a 14.5 - 15.5 Hz filtered Gaussian noise carrier.
Test signals were simulated with a duration of 1000 s sampled at 100 Hz and embedded in 6 dB additive Gaussian white noise. HOSD was carried out with the trispectrum windowed within the diagonal slice. 
}
\end{figure}

\section*{Materials and Methods}

\subsection*{Data segmentation}
Following line noise removal and artifact rejection \cite{kovach2016demodulated}, the analysis of discretely sampled data begins by dividing the signal into $J$ overlapping intervals of duration $P$ samples each, tapering each interval with a window function $w[n]$ to reduce spectral leakage, and computing the discrete Fourier transform (DFT) of each interval:
\begin{align}
\label{EQ:DFT}
\begin{split}
X_j[m] = \sum_{p=0}^{P-1} x[p+\Delta_j]w[p]e^{-i\frac{2\pi p}{P} m }
\end{split}
\end{align}
Segments were tapered with Sasaki's window function \cite{sasaki1975minimum}.

\subsubsection*{Window selection}
A common approach to selecting the window delays, $\{\Delta_j\}$ in (\ref{EQ:DFT}) maintains a fixed overlap between successive window, typically 50\% of the window support.
When the analysis concerns particular events of interest, \emph{event related windowing}, selects $\{\Delta_j\}$ to center the analysis windows of event times. 
Windows selected in this way may be understood as recovering a degenerate subdomain of the $5^\text{th}$-order cross-moment spectrum between the series of impulses representing event times and the signal. The proper computation of the cumulant spectrum should therefore account for lower-order terms;  for simplicity, event-related analyses described later represent moment spectra rather than cumulants.

\subsection*{Modulogram estimation}
The modulogram is equivalent to the normalized direct estimator of the trispectrum within the diagonal slice.
For each tapered segment, the discrete Fourier transform was obtained according to \refeqn{EQ:DFT}. 
Coefficients were estimated according to \refeqn{EQ:directHOS} at each frequency triple within the principal diagonal slice by averaging the corresponding product of Fourier coefficients across segments:
\begin{align}
\begin{split}
\label{EQ:directMds}
\widehat{\mathrm{MG}}[k,l]  = \hat{M}_{DS}^4[k,l] &= \frac{1}{J}\sum_{j=1}^J X^2_j[k]X^*_j[k-l]X^*_j[k+l]
\end{split}
\end{align}
Tricoherence was obtained by normalizing \refeqn{EQ:directMds} by the mean magnitude of the products according to \refeqn{EQ:discrete_polycoh}.  

\subsection*{HOSD implementation}
HOSD was applied to the trispectrum windowed within the diagonal slice.
Details of this procedure are given in Appendix \ref{SEC:HOS_METHODS}.


\subsection*{Kurtogram}
The kurtogram was computed according to \cite{antoni2007fast} using the function, KURTOGRAM, implemented in the Matlab (ver. R2021a) signal processing toolbox.
It represents spectral kurtosis (SK) computed with analysis windows of varying durlation.
SK was computed using windowed spectral power obtained from (\ref{EQ:DFT}) at multiple window sizes, $P$, as the kurtosis of the squared norm of the DFT across analysis windows:
\begin{align}
\label{EQ:SK}
SK _P[m]= \frac{J(J+1)}{J-1}\frac{\sum{\left|X_j[m]\right|^4}}{\left(\sum{\left|X_j[m]\right|^2}\right)^2}-2
\end{align}
This effectively represents a normalized estimate of the degenerate region of the trispectrum given in \refeqn{EQ:trispect_degen}, with $\omega_1=\omega_2$.

\subsection*{Continuous wavelet analysis}
Time frequency plots used in {\bf fig. \ref{fig:HOSD_betaburst_example_R1S1}} applied a Morlet wavelet transform with bandwidth parameter, $\sigma=3$.  

\subsection*{Simulations}
Simulated data were generated by multiplying a Gaussian carrier signal with a non-negative modulating signal.  
The Gaussian carrier was obtained by filtering Gaussian white noise within a specified frequency band. 
The modulating signal was obtained by passing Gaussian noise, filtered within a lower frequency range and standardized, through the logistic sigmoid nonlinearity with scale parameter, $\sigma=2$.
Further details are provided in the caption of {\bf figure \ref{fig:mdg_ktg_comp} }.  

\subsection*{Beta Burst Detection in Rodent LFP Data}

\subsubsection*{Data set}

Beta burst detection was demonstrated using an open dataset \cite{KarvatRepo} of single-channel laminar field potential recordings (LFP) targeted to layer 5 of the rat motor cortex.
The original study of Karvat et al. examined whether beta (12-40 Hz) power was altered by a conditional reward of sucrose water, delivered whenever power exceeded an adaptive threshold \cite{karvat2020real}.
Putative burst events were detected in the study using realtime processing and reward delivery.
Over the course of  9 sessions, lasting 30 minutes each, in each of 3 rats, power within a selected band was adaptively thresholded such that approximately 100 rewards were delivered over a session. 
For details on recordings and conditioning parameters, see Karvat et al. 2020. 

\subsubsection*{Modulogram estimation}
Data were first downsampled to 244.14 Hz from 976.56 Hz then divided into 6 second (1465 sample) intervals and tapered with a Hann window. 
Modulograms in {\bf Fig. \ref{fig:BB_session_comparison}A} were computed separately for each of 9 session.
Analysis windows were chosen in two way; first, in an event-focused manner by including windows centered on the reward times, thus emphasizing features associated with transient increases in beta power as detected in the realtime analysis of Karvat et. al. 
The second method divided the full recording into overlapping 4 second window with 50\% overlap and was therefore fully blind to experimental procedure.
The second approach should be free of any potential bias that might arise from selecting windows according to the realtime analysis of Karvat et al, while the first may be expected to improve sensitivity to reward-associated events of interest.
A third analysis computed the modulogram on data concatenated over all 9 session ({\bf Fig. \ref{fig:BB_session_comparison}B}), using event-independent segmentation.
\subsubsection*{Event identification}
\begin{figure*} 
\includegraphics[width=\textwidth]{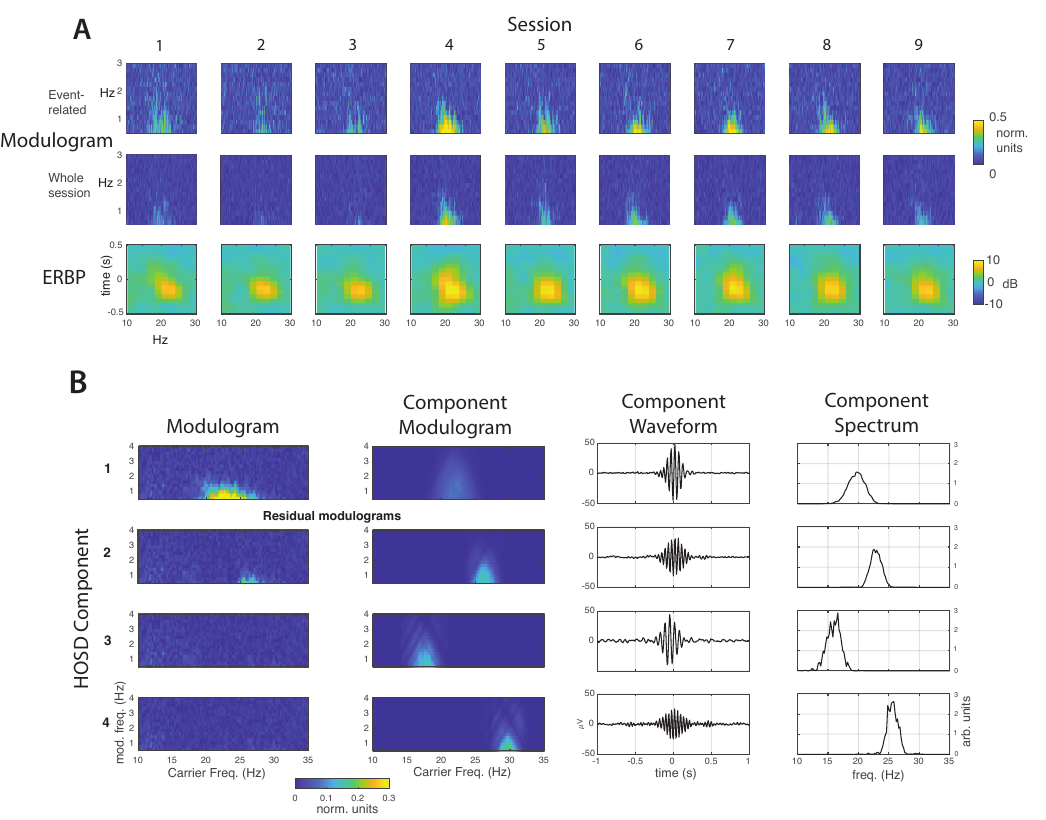}
\caption{\label{fig:BB_session_comparison} 
Beta burst identification with the modulogram in rodent LFP recording, demonstrated with data from an open data set shared by Karvat et al. (2020). 
{\bf A}: Modulograms were computed for data from a single recording channel over 9 sessions from rat 1 using two approaches. 
In the first case, analysis windows were selected to include the reward delivery period (\emph{A, top row}).
In the protocol of Karvat et al, the reward is conditioned on the occurrence of peaks in beta power. This introduces the possibility of bias in the reward-centered analysis, thus in the second approach (\emph{A, second row}) analysis windows were drawn from the entire recording session with 50\% overlap.
Both methods show a clear band of modulation with a carrier frequency range from 19 to 25 Hz, being somewhat more prominent in the event-centered windowing.
Spectral characteristics of amplitude modulation are visible along the vertical dimension, with energy extending from $<0.5$ Hz to about $1.5$ Hz, consistent with aperiodic burst-like envelope modulation.
Mean peri-reward time-frequency power is shown for comparison (\emph{A, third row}); note the peak in power is consistent with the association of reward with peaks in beta power detected online.
{\bf B}:  Component recovery with HOSD, using data concatenated across sessions. 
The first component recovered with HOSD explains the central portion of the AM band in the modulogram (\emph{B, top row, second column}).
Subsequent components (\emph{B, rows 2 - 4}) capture residual energy within the upper and lower ranges of the carrier band, most likely related to variability of the burst frequency.
The recovered waveform associated with the $1^\mathrm{st}$ component (\emph{B, top row, third column}) reveals the characteristic duration and form of burst-related amplitude modulation.
Because of its nature as a projection into a lower-dimensional feature space, HOSD is able to recover oscillatory components even when the component is not obvious in the modulogram magnitude plot. 
}
\end{figure*}

To compare oscillatory bursting across sessions HOSD \cite{Kovach2019} was carried out on data concatenated across sessions.
Concatenated data were used in this analysis to facilitate comparison across sessions using a common set of features, as HOSD might otherwise return different features when applied separately to each session. 
HOSD was applied to the trispectral diagonal slice, windowed to include carrier frequencies between 10 Hz and 40 Hz and envelope modulation frequencies between 0.2 Hz and 5 Hz. 
These spectral ranges entailed the estimation of 5,636 coefficients at the frequency resolution afforded by the 5 s analysis window. 
A reconstructed signal was obtained by sequentially estimating and summing component signals until the excess kurtosis of the filter output at a given component fell below 0.1. 
Example component waveforms and associated modulograms are shown for rat 1 in {\bf Fig. \ref{fig:BB_session_comparison}B}.
To identify individual burst events, peak detection was applied to the out put of each recovered HOSD feature detection filter, thresholded such that the excess kurtosis of the residual (subthreshold) signal was 0 then smoothed with a 0.1 s (24 sample) duration Hann window.

\subsubsection*{Event characterization}

 The initial time estimates were adjusted by reassigning event times to the nearest peak within the analytic envelope of the summed reconstructed signal, with a tolerance of $\pm$0.5 s around the preliminary estimate. 
Because the HOSD detection filters are non-zero phase, the latter step helped to ensure that the final event times aligned better with peaks in band limited power as observed using a standard minimum phase IIR or zero-phase FIR filter, with minimal phase-related systematic offset.

Following this procedure, burst amplitudes were identified as the magnitude of the analytic envelope of the reconstructed signal, summed over components, at the respective final event times. Event center frequency was identified by computing analytic instantaneous frequency from the phase derivative at associated peak times in component filter outputs, then computing a weighted average across components, in which instantaneous power at the peak time served as the weighting factor.  

\begin{figure*} 
\includegraphics[width=\textwidth]{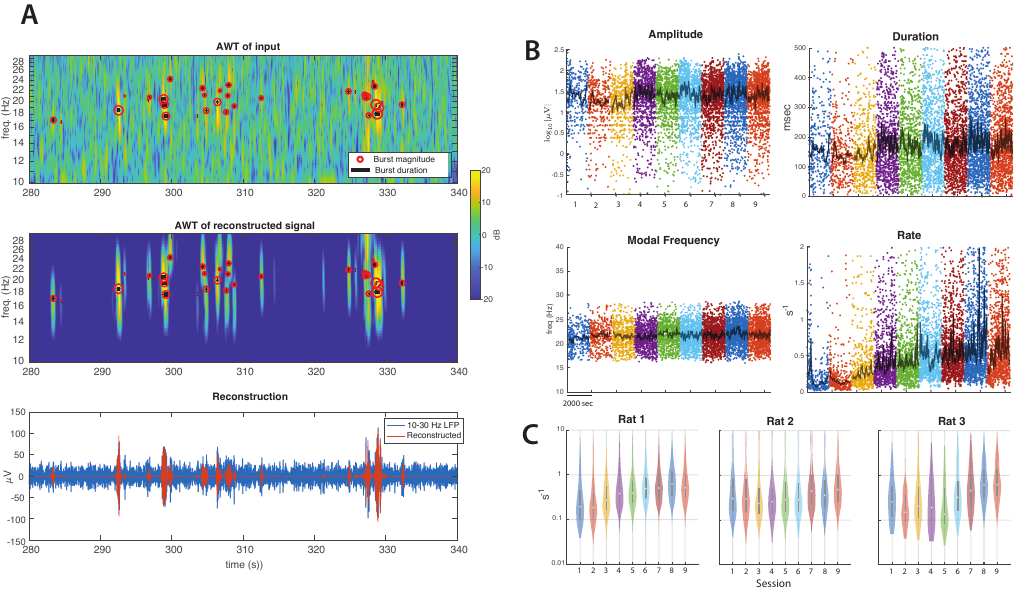}
\caption{\label{fig:HOSD_betaburst_example_R1S1} 
{\bf A}: Identification of beta bursts from HOSD components. 
Beta bursts identified according to above-threshold peaks in the root-mean-square (RMS) smoothed filter outputs with reconstructed components from HOSD; an example is shown for rat 1, superimposed on the wavelet scalogram of the LFP signal (\emph{A, top panel}) and on the that of the reconstructed signal summed over the first four HOSD components (\emph{B, middle panel}).
Burst times and magnitudes were determined according to peaks in the analytic amplitude of the reconstructed signal, shown in the bottom panel (\emph{A, red trace}).
The center frequency and duration of each burst were taken, respectively, from instantaneous frequency and the full width at half maximum (FWHM) duration at respective peak times.
Note that burst detection is not derived from the displayed wavelet decompositions, which are shown for comparison. 
{\bf B}: Characteristics of beta burst detected with HOSD over time for rat 1: amplitude (\emph{B, top left}), duration (\emph{B, top right}), rate (\emph{B, bottom right}) and instantaneous frequency at peak amplitude (\emph{B, bottom left}). 
Lines show median values within a sliding window of 50 data points.
Changes of over time and session are most evident for rate. 
{\bf C}: In all 3 rats, burst rate varied over sessions. 
Violin plots show the distribution (\emph{shaded}), median (\emph{circle}) and interquartile range (\emph{gray line}) for burst rate, computed as log inverse inter-burst interval, which trended upwards over the course of reward conditioning. Included in the analysis are HOSD components with $\ge$ -20 dB median SNR in the 12-40 Hz band. 
}
\end{figure*}

 Burst duration was measured as the full-width-at-half-maximum (FWHM) interval of the analytic envelope of the reconstructed signal with respect to the peak magnitude.
 An example of detected events from session 1 of rat 1, comparing the original and reconstructed signal, is shown in {\bf Fig. \ref{fig:HOSD_betaburst_example_R1S1}}.

%
%

\begin{figure*}
\includegraphics[width=.66\textwidth]{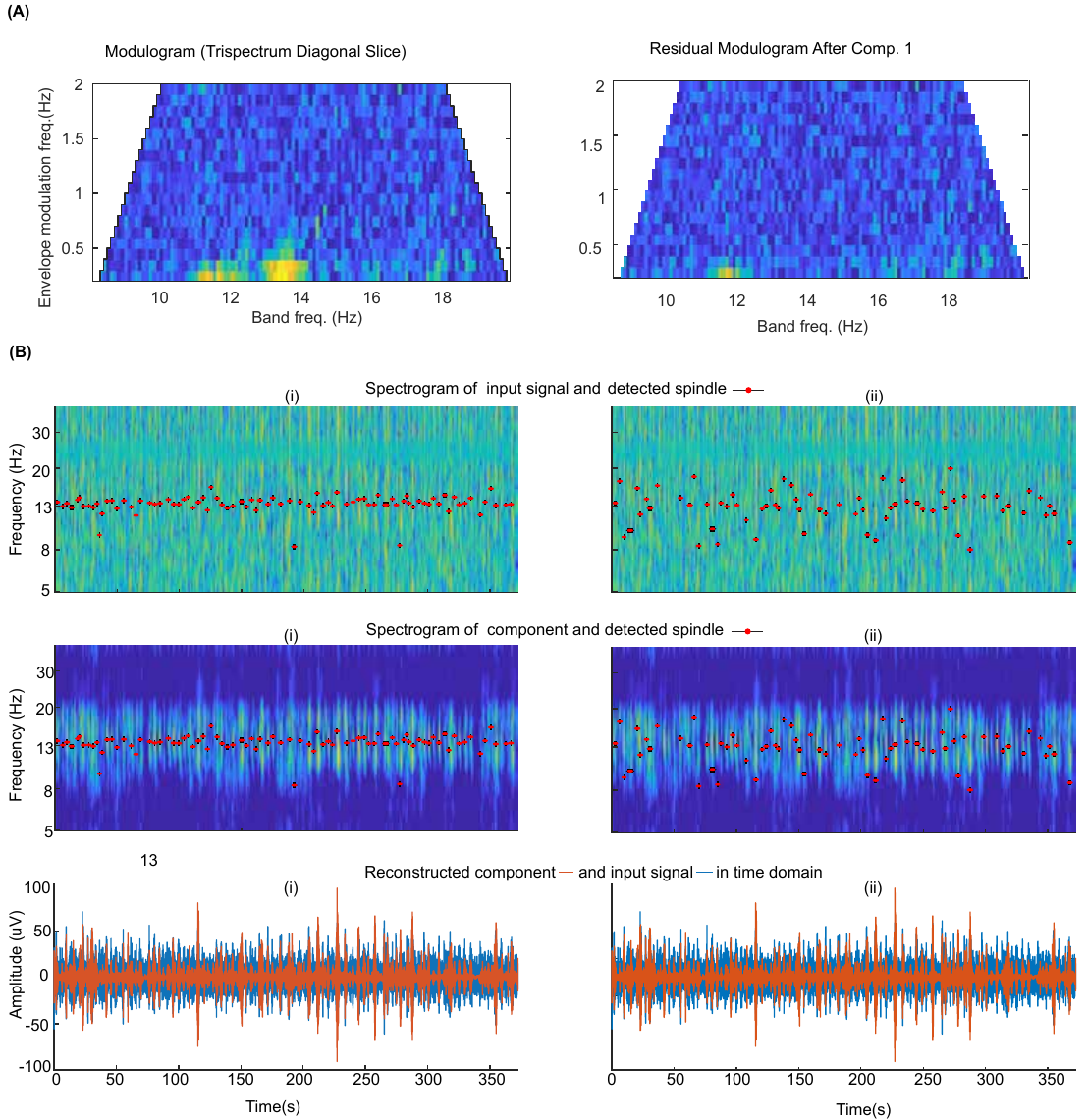}
\caption{\label{fig:Spindle}
Detection of spindles using the diagonal slice of the trispectrum.
 The trispectrum was estimated using carrier frequencies between 8 and 20 Hz and temporal envelope frequencies between 0.1 and 2 Hz. 
 Spindle timing and amplitudes were determined based on the local maxima of the analytic amplitude of the reconstructed signal. 
\emph{Top}: Detected spindles superimposed on the time-frequency spectrogram estimated via wavelet transform of the raw signal. 
\emph{Middle}: Detected spindles overlaid on the time-frequency spectrogram of the reconstructed first higher-order spectral decomposition (HOSD) component. \emph{Bottom}: Time-domain plot of the reconstructed first component overlaid on the raw signal.
}
\end{figure*}

\subsection*{Spindle Detection}
Sleep spindles are hallmark oscillatory bursts observed during non-rapid eye movement (NREM) sleep, characterized by a distinct frequency range of 11 to 16 Hz and a duration lasting at least 0.5 seconds \cite{fernandez2020sleep,schonauer2018sleep}.
Notably, these oscillatory bursts are not exclusive to natural sleep; spindle-like bursts are also prominently observed under general anesthesia. 
To validate our method, we demonstrate the application of the trispectrum for automatic identification of spindles using data from one human subject with intractable epilepsy who was implanted with intracranial electrodes for seizure focus localization. 
Specifically, we analyzed local field potential (LFP) recording obtained from the thalamus in a subject sedated with propofol.

\subsubsection*{Data set}
The subject was a 39-year-old patient with medically intractable epilepsy who underwent stereoencephalography (sEEG) and was monitored over a two-week period via intracranial electrode recordings. 
The electrodes (Ad-Tech Medical Instrument Corporation, Oak Creek, WI) were implanted at sites determined by the clinical epilepsy team at the University of Iowa.
Approximately 7 minutes of LFP data were recorded prior to electrode explanation following anesthesia induction. 
The data were sampled at 2000 Hz and recorded from a site located in the thalamus while the subject was sedated under propofol anesthesia and breathing spontaneously. 
The experimental protocol was approved by the University of Iowa Institutional Review Board.

The LFP data were downsampled to 250 Hz, divided into 10-second epochs, with each tapered by a Hann window. 
Recurrent features were then isolated by applying HOSD to the trispectral diagonal slice. 
The analysis followed the oscillatory burst detection procedure described above, modified to include carrier frequencies between 8 and 20 Hz, centered on the spindle band,  with envelope modulation frequencies between 0.1 and 2 Hz.
{\bf Figure \ref{fig:Spindle}} shows the modulogram and detected oscillatory bursts from the first two components returned by HOSD.

\section*{Results}

\subsection*{AM Detection in Simulated Data}
The detection of narrowband amplitude modulation of a band limited carrier, using the modulogram, is illustrated in {\bf Figure \ref{fig:mdg_ktg_comp} }.
The duration of the simulated data, 1000 s, and signal to noise ratio, $-3.5$ dB (noise amplitude $=1.5\;\times\;$signal amplitude), were chosen to be plausible for typical experimental recordings of local field potentials.  
Comparing the modulogram to the kurtogram ({\bf Fig. \ref{fig:mdg_ktg_comp}}, \emph{third row}) illustrates the superior performance of the former in resolving the spectrum of the modulating signal. 
The kurtogram is able to detect amplitude modulation only using analysis windows shorter than the time scale of modulation, which sacrifices both frequency resolution and robustness to noise.
At low fractional bandwidths, the kurtogram fails to clearly reveal amplitude modulation altogether in the presence of noise.
Outside of the DC band, the modulogram detected AM within the opposite range of time scales, shorter than the analysis window. 

\subsection*{Distinguishing AM and FM}
Amplitude and frequency modulation both induce correlations between upper and lower sidebands, but with opposite sign: the correlation is positive for AM and negative for FM. 
This distinction is plainly evident in modulogram phase, as shown in {\bf Figure \ref{fig:am_vs_fm}}, which is $0$ rad for a simulated AM signal and $\pi$ rad for a FM. 
A time-frequency plot of the first feature returned by HOSD, applied to respective signals, shows the characteristic difference between amplitude and phase modulation ({\bf Fig. \ref{fig:am_vs_fm}}, \emph{bottom row}).
Intermediate phases convey the degree to which amplitude in the upper sideband leads the lower side band, illustrating how modulogram phase may encode spectrotemporal patterns of modulation more generally. 

\subsection*{Beta Burst Detection}
 
Applications of the modulogram and HOSD are demonstrated over a range of signal to noise ratios with LFP recordings from rodent motor cortex published by Karvat et al.  \cite{KarvatRepo}.  
Details of the experimental procedure are given in \cite{karvat2020real}.
In brief, data were obtained during a reward conditioning paradigm in which a reward of sugar water was delivered following increases in power within the beta band, detected in realtime.

For the first subject (rat 1),  the modulogram revealed clear modulation in the beta range in all 9 sessions, using both reward-centered and reward-independent windowing ({\bf Figure \ref{fig:BB_session_comparison}A}).
Comparison of modulograms computed with event-related windowing, centered on reward events, and event-independent windowing shows somewhat greater relative enhancement in the former.
Both methods yield evidence of modulation in the beta range.
In both cases, peak carrier frequencies align well with the frequencies on which rewards were contingent, as detected in the real-time analysis of Karvat et al.

The application of HOSD towards isolating beta-burst-associated features and events is illustrated in {\bf Figures \ref{fig:BB_session_comparison}B}, {\bf \ref{fig:HOSD_rat2_and_3}} and {\bf \ref{fig:HOSD_betaburst_example_R1S1} }. 
For rat 1, the first component returned by HOSD is centered on the band of modulation evident in the modulogram, while subsequent components explain energy at the periphery, likely reflecting variability of the center frequency across bursts  ({\bf Fig. \ref{fig:BB_session_comparison}B}).	

The same analysis was repeated with data from two additional subjects (rats 2 and 3).
The event-independent modulogram in rat 2 produced faint evidence of oscillatory bursting in beta and alpha bands and no obvious signature of oscillatory bursting in rat 3 ({\bf Fig. \ref{fig:HOSD_rat2_and_3}}).
In spite of these seemingly equivocal findings, HOSD recovered oscillatory components over a range of frequencies for both rats, which included the beta band. 
Moreover, these responses exhibited behavioral modulation in agreement with the findings of Karvat et al. ({\bf Fig. \ref{fig:HOSD_betaburst_example_R1S1}}C).

The detection and characterization of individual bursts using features in the beta range, identified by HOSD, is illustrated in {\bf Figure \ref{fig:HOSD_betaburst_example_R1S1}A}.
Duration, center frequency, amplitude of individual bursts and rate of burst events are shown for rat 1 in {\bf Figure  \ref{fig:HOSD_betaburst_example_R1S1}B}. 
Violin plots show the distribution of event rates for each session in all three rats  in {\bf Figure  \ref{fig:HOSD_betaburst_example_R1S1}C}.
In all three cases, burst rate increased over the course of training, in agreement with the observations of Karvat et al. \cite{karvat2020real}. 

For insight into performance in detecting weak modulation in noise, signal-to-noise ratio (SNR) was estimated within a 200 ms window around identified burst times according to the ratio of mean power in the reconstructed signal and the residual. 
For rat 1, median overall SNR at burst times, within the frequency range 10 to 40 Hz, was -9.4 dB, with inter-quartile range of -22 dB and -0.7 dB. 
SNR varied by HOSD component and generally decreased with component number.
The first component had greatest overall estimated SNR with median -4.8 dB and inter-quartile range of -12 dB to 2.6  dB. 
Estimated SNR was markedly lower for rats 2 and 3:  for rat 2, the median for the component with greatest SNR in the beta range as -9.1 dB with inter-quartile range -22 dB to -0.24 dB, while for rat 3, the median was -17 dB with inter-quartile range -24 dB to -11 dB.

\subsection*{Spindle detection}
The current method was also demonstrated using a brief (approximately 7 min.) recording of human intracranial data obtained from the thalamus during anesthesia induction, which contained spindle-like oscillatory bursts.   
As shown in  {\bf Figure \ref{fig:Spindle}}, HOSD recovered two spindle-like components, the first with a frequency around 14 Hz, and the second around 12 Hz. 
The modulogram reveals the characteristic frequencies of both bands as well as the slower time scale of modulation, compared to beta bursts.

\section*{Discussion}

This work has sought to demonstrate the usefulness of the trispectrum for identifying forms of modulation, beginning with the derivation of the trispectrum as a measure of linear relationships of power across frequencies, as expressed by \refeqn{EQ:MSmom}.
This relationship is useful for building intuition about the meaning of the trispectrum, and may help to overcome a supposed obscurity of interpretation as a major obstacle to the wider adoption of HOS-based methods.
Although it is rather simple to demonstrate, the author is unaware of previous work to have made use of the relationship.

Beyond the question of interpretation, trispectral estimators avoid important deficiencies of techniques that begin with the computation of time-varying power using conventional methods of filtering and rectification, short-time Fourier transforms, or equivalents.
These include the unavoidable biases arising from the uncertainty relationship in extracting power from filtered data, as well as biases that result from squaring additive Gaussian noise.
Higher-order spectra, like the trispectrum, are inherently robust to interference from additive Gaussian noise and allow for asymptotically unbiased estimation, free of the assumptions that must otherwise go into the preliminary step of computing time-varying power.  

Preservation of phase is another key advantage of HOS-based techniques, which allows them to retain detailed information about characteristic spectrotemporal patterns of modulation. 
This information allows different types of modulation to be distinguished  ({\bf Fig. \ref{fig:am_vs_fm}}) or recurring patterns to be identified.
The need for methods that capture details of wave shape, lost in standard spectral analyses, has come increasingly into focus \cite{cole2017brain}.
By preserving phase, HOS offer a comprehensive and mathematically principled set of tools for addressing this problem.
Moreover in the context of HOS, ``wave shape'' takes on a broad meaning, as it might refer to characteristic patterns at the first order (i.e. waveforms in the raw signal) as well as higher orders, exemplified here with second-order nonstationarity in the form of envelope modulation. The trispectrum retains information needed to identify the shape of second-order nonstationarity, as well as first-order. 

A drawback of HOS is their high dimensionality and the correspondingly large number of coefficients that may need to be estimated. 
As a three-dimensional function, whose dimensionality grows with the cube of the number of samples in the analysis window, the direct computation, representation and interpretation of the trispectrum may be especially cumbersome. 
The two-dimensional diagonal slice in \refeqn{EQ:DSdef}, however, represents essential spectral properties of modulation in an interpretable form, provided that energy is preserved in the carrier band, which requires the envelope to be lowpass or to have a non-zero DC offset. 

For processes that may not meet these requirements, the scheme can be modified; for example integrating over $\eta$ in \refeqn{EQ:MSmomreparam} gives the zero-lag correlation between sidebands, regardless of whether the carrier is preserved by modulation.
In general, developing a framework for interpreting HOS allows the dimensionality problem to be attacked through selective windowing, according to the relevance of particular subdomains for a given question.
The present application shows how to guide such a framework by understanding how HOS relate to dynamic properties of lower-order statistics.

The decomposition of HOS into additive deterministic spectra carries the resolution to problems of dimensionality and interpretation still further. 
HOSD provides a form of dimensionality reduction in which each component is represented by a concrete and interpretable waveform, whose locations can subsequently be identified in original data. 
As illustrated with oscillatory burst identification in rats 2 and 3, such a projection may serve to squeeze additional information from the trispectrum by recovering structure otherwise obscured by noise in the direct representation of spectral coefficients ({\bf Fig. \ref{fig:HOSD_rat2_and_3}}).


\subsection*{Relationship to ICA}

Independent component analysis (ICA) seeks to separate statistically independent signals from an unknown linear mixture within a multivariate signal. 
ICA operates through a decomposition of the higher moments of a multivariate process \cite{cardoso1990eigen,cardoso1991super,comon1994independent,bell1995information,hyvarinen1999}, which often remain implicit within the polynomial expansion of some static nonlinearity.
In the same way that principal component analysis (PCA) obtains a series of projections that sequentially maximize variance, ICA finds a series of projections that maximize some combination of static moments.
Higher moments, computed directly or implicitly, tend to provide a more effective criterion for the separation of independent processes than does covariance as used by PCA. 
Separating otherwise correlated features is enabled by the high dimensionality of the space such moments inhabit and by the additive property of cumulants for independent processes. 

Because it also yields an additive decomposition, HOSD can be understood as a type of statio-temporal
ICA, applicable when sources outnumber channels  \cite{cardoso1991super}.
But as with ICA generally, identifying components with independent sources should be approached tentatively, with underlying assumptions kept in mind.
HOSD effectively relies on direct cumulant maximization, making it appropriate for separating super-Gaussian sources, but not sub-Gaussian.
Components may also reflect (higher-order) non-stationarity in a single source rather than a mixture of independent sources; this phenomenon seems likely to explain the presence of multiple beta-band components in the rodent data, arising from variability of burst frequency rather than independent sources with different fixed frequencies. 
How best to distinguish these cases remains an open problem.
Even so, the resulting projection into a low dimensional feature space remains a useful aid to interpretation and analysis.


\subsection*{Relationship to Blind deconvolution}
Blind deconvolution seeks to recover a signal that has been distorted by a linear system. 
Common techniques work by finding a linear filter that optimizes a static non-linearity \cite{wiggins1978,comon1994independent,cadzow1996}.
They can be motivated by arguments very similar to those for ICA \cite{bell1995information}, with the constraint that mixing (over time rather than channels) is described by a Toeplitz matrix or, in the case of FIR estimation, banded-Toeplitz, which implies finite memory \cite{comon1994independent}.
The banded Toeplitz constraint is implicit in the use of time-shift-invariant HOS estimated with short duration analysis windows.  

Classical techniques of blind deconvolution maximize the diagonal of a given static cumulant directly \cite{wiggins1978,donoho1981minimum,mcdonald2012maximum,ovacikli2016recovering}, rather than the magnitude of the diagonal. This assumes the sources to be be super-Gaussian, making the algorithms unsuitable for the separation of sub-Gaussian sources (i.e. with negative excess kurtosis) \cite{bell1995information}, although algorithms suitable for both cases also exist \cite{tugnait1997identification}.
In signal processing contexts, the assumption that the target signal is super-Gaussian often makes sense. 
For example, when the goal is to detect some randomly recurring feature, an appropriate model might be a linear system which convolves the feature waveform with the output of an impulse generating process, responsible for the feature emission times.
In the present setting, amplitude modulation of a Gaussian carrier results in super-Gaussian statistics \cite{antoni2006spectral}, thus the assumption in an appropriate one.
More generally, white noise modeled as an infinitely divisible independent increment process, also known as a L\'evy process, can be decomposed into, at most, a continuous Gaussian process and a discontinuous ``jump'' process, with statistics of the latter strictly super-Gaussian \cite{applebaum2009levy}. 
For this reason, no process driven by continuous-time white noise can be sub-Gaussian.

\subsection*{Relationship to Cyclostationarity}
The basic point that $2^\text{nd}$-order nonstationarity is reflected in specific kinds of cross-frequency correlation has been considered in detail for \emph{cyclostationary} (CS) processes, whose nonstationary $2^\text{nd}$-order statistics vary with fixed periodicity \cite{gardner1991exploitation}. 
A quantity similar to the modulation spectrum \refeqn{EQ:MSDdef} is called in this literature ``spectral correlation density'' (SCD) \cite{gardner1991exploitation}. 
The strictly periodic nature of cyclostationary systems means that even at low signal to noise ratios, the system may often be identified through the application of a quadratic or similar nonlinearity, which yields back the underlying periodicity (e.g. periodic amplitude modulation) in the power spectrum of the result. 
The potential difficulty caused by squaring Gaussian noise, considered earlier, is often negligible for cyclostationary systems because the large peaks, or ``spectral lines,'' that emerge in the power spectrum of the transformed signal stand out sharply from spectrally diffuse additive noise.
Thus, for the purpose of detecting a CS process, there is often no need to consider anything more sophisticated than peak detection within the power spectrum of the transformed output.
This consideration plainly fails to apply to the general case that interests us, and cumulant properties of HOS become of great value in separating an aperiodic or weakly periodic signal from Gaussian noise.
Because the trispectrum contains the complete cross spectrum for bands within the WVD, it conveys information about dynamic properties of aperiodic as well as periodic amplitude modulation, and does so without bias from additive Gaussian noise.



\subsection*{Relationship to Spectral Kurtosis and the Kurtogram}
Spectrally resolved kurtosis as a measure of $2^\text{nd}$-order nonstationarity has been considered by a number of authors \cite{dwyer1984use, capdevielle1996blind,vrabie2003spectral}. 
A quantity, known as spectral kurtosis (SK), is defined (most commonly) as kurtosis computed over the time dimension of a short-time Fourier transform \cite{dwyer1984use} as in \refeqn{EQ:SK}.
Capdevielle considered the relationship of SK to the normalized trispectrum \cite{capdevielle1996blind} within what is here considered the 0 Hz modulation band of the diagonal slice, as given in \refeqn{EQ:DSdef}.   
Antoni raised the critique that identifying SK with a subdomain of the trispectrum leads to difficulty in applying the concept to nonstationary signals  \cite{antoni2006spectral}, arguing that the use of time-invariant statistics contradicts the goal of describing non-stationarity in the signal.
The present discussion should clarify the source of the conceptual difficulty, which is not in the application of the trispectrum to (conditionally) nonstationary signals, as such.
Rather, an internal contradiction arises from the confinement of SK to the 0 Hz modulation band in what is here identified as the trispectral diagonal slice.
This seems conceptually to limit SK to signals with no amplitude modulation.
In practice, the SK estimator relies on windowing-related summation of energy around the DC band of the modulating signal as a consequence of broadband bias within the estimator.
For this reason, SK conveys excess kurtosis only at time scales longer than the analysis window and fails if the window duration does not match the characteristic time scale of any amplitude modulation.

Antoni proposed to address these shortcomings by computing SK at multiple time scales \cite{antoni2006spectral}, yielding a two dimensional function, the ``kurtogram,'' with frequency varying along one axis and window scale along the other \cite{antoni2006spectral2}. 
Like the modulogram, the kurtogram conveys information about both the carrier band and time scale of characteristic amplitude modulation. 
From the trispectral standpoint, however, it still represents a smoothing of the trispectrum within the DC band under varying levels of broadband bias, which falls short of completely addressing the conceptual shortcomings of SK. 
For the same reason, the kurtogram is severely limited in its ability to resolve spectral properties of the modulating signal ({\bf Fig. \ref{fig:mdg_ktg_comp}}). A further consequence is that it cannot detect frequency modulation, because the anti-correlated amplitudes in the sidebands of an FM signal cancel upon being blended together by broadband bias. 

\subsection*{On the modulation spectrum}
The magnitude of the modulation spectrum, as given in \refeqn{EQ:MSDdef}, has been been proposed as a tool for identifying modulation in its own right  in the context of audio \cite{atlas2003joint} and cyclostationary \cite{gardner1991exploitation} signals. 
The expectation of the squared modulation spectrum yields a degenerate subdomain of the 4$^\text{th}$-order moment spectrum.
There is however an important flaw in the use of the modulation spectrum for identifying AM. 
As noted earlier, on its own, it does not appropriately convey the correlation between sidebands that is the hallmark of AM and FM, while its expectation is biased by the power spectrum of the signal. 
For example, it is visually evident in {\bf Fig. \ref{fig:wvd_examp_random}} (\emph{2$^\text{nd}$ row}), that the spectrum of the WVD in the phase randomized example  must contain a bump midway between the two (independent) side-bands at the same modulation frequency as the AM signal ({\bf Fig. \ref{fig:wvd_examp_random}}, \emph{top row}), even though there is no modulation in the former. 
This is a consequence of the fact that for a stationary Gaussian signal (which has no modulation) the moment spectrum does not vanish:
\begin{align}
\label{EQ:MSEGaussian}
\begin{split}
&\expect{\left|\mathcal{M}_\mathrm{Gauss.}\left\{x\right\}(\omega,\eta)\right|^2}  =\\
&\quad\expect{\left|X\left(\omega+\frac{\eta}{2}\right)\right|^2}\expect{\left|X^*\left(\omega-\frac{\eta}{2}\right)\right|^2}
\end{split}
\end{align}
The modulation spectrum, by itself, does not account for this bias.
Because it is not defined as an expectation, it retains the ambiguity of the WVD with respect to the signal model and cannot be adjusted to account for bias.

\subsection*{Identifying oscillatory bursts with the modulogram and HOSD}

Oscillatory bursting in electrophysiological recordings is an area of burgeoning interest \cite{wessel2020beta}, yet the literature still lacks a clear set of coherent and principled criteria for identifying its presence, magnitude and fundamental properties.
The ability of the modulogram to fulfill this role is demonstrated with local field potential recordings from one subject (rat 1) in the published data set of Karvat et al. (2020), summarized in {\bf Figs. \ref{fig:BB_session_comparison}  and \ref{fig:HOSD_betaburst_example_R1S1}}.
For rat 1, bursting manifests as low frequency modulation within the beta band, observed in both event-related and event-independent analyses, where the event was delivery of a food reward.
The evidence for bursting from the modulogram, alone, is considerably weaker for rat 2 and altogether absent in rat 3, likely reflecting lower signal-to-noise ratios in these recordings. 
Nevertheless, HOSD recovered oscillatory components within the target frequency range, demonstrating its ability to boost sensitivity to oscillatory features through the projection into a lower-dimensional feature space.  
A rough guide to the sensitivity of the blind method used here can be obtained by comparing relative power in the reconstructed and residual signals.
For rat 1, in the 12 Hz to 40 Hz band, the highest median SNR across all recovered components, computed 200 ms around burst times, was -4.8 dB,  normalized to the residual signal.
Maximum estimated SNRs for rats 2 and 3 were markedly lower at  -9 dB and  -17 dB, respectively. 
These values provide a rough boundary on the relative magnitude of oscillatory bursts, warranting some confidence that the algorithms perform satisfactorily even in the context of low SNRs.

Karvat et al. reported increased beta-band power in all 3 subjects following conditioning to a liquid reward, consistent with changes of bursting rate observed here in the same data ({\bf Fig. \ref{fig:HOSD_betaburst_example_R1S1}C}).
That the most prominent change observed here is burst rate, not amplitude or duration, agrees with with previous reports on characteristics of beta bursting \cite{shin2017rate}.
Thus, in spite of the failure of the modulogram to produce unequivocal evidence for bursting in data from rats 2 and 3, oscillatory features identified with HOSD have expected characteristics for beta bursts.

Because HOSD relies on the same information represented in the modulogram, this apparent discrepancy in the sensitivity of the methods deserves some scrutiny. 
Coefficients within the trispectral estimate are asymptotically independent \cite{brillinger1967asymptotic}, so not restricted by any assumed structure within the data. 
The dimensionality of the trispectral estimate thus asymptotically equals the number of coefficients estimated within the principal part of the spectrum. 
HOSD, on the other hand, takes advantage of structure within HOS to obtain a low dimensional projection onto components whose separate HOS have  deterministic structure, meaning they obey the expected product relationship; in this case $M^4_\mathrm{DS}(\omega_1,\omega_2) = F(\omega_1)F(\omega_2)F^2\left(-\frac{\omega_1+\omega_2}{2}\right)$.
The dimensionality of the manifold implied by this structure is equal to the number of non-zeros frequency samples in F, which, in general, is substantially lower than the dimensionality of the HOS estimate. 
The greater apparent sensitivity of HOSD  in revealing structure within the data, compared to raw HOS estimates, may therefore be attributed the implicit projection into a lower-dimensional manifold with the structure of a deterministic spectrum.

\section*{Conclusion}

Linking blind identification to specific and interpretable representations of the underlying higher-order statistics is an important aspect of the present work.
 In spite of the considerable recent progress in computational methods for identifying and decomposing signals according to a variety of criteria \cite{wang2012nonnegative, sidiropoulos2017tensor}, a need for a general, comprehensive and unbiased framework for identifying and separating non-Gaussian signal features persists. 
Techniques using higher-order moments have a long history of fulfilling this need in many settings, with especially well established applications to blind source separation for multivariate signals.
Nevertheless, such methods collectively tend to suffer from a ``black box'' problem by offering little opportunity for \emph{a-priori} insight into the features and properties of a signal that cause a given algorithm to converge on a given solution.
For this reason, ICA and related techniques find the widest use in the separation of signals that are at least partially known, for which success is easy to verify, such as EKG, and for nuisance signals, such as artifactual or physiological interference in EEG, for which the nature of the source is less important than the practical end of scrubbing it from the signal. 
As a tool for gaining scientific insight, such methods are still held back by questions of interpretation and reliability \cite{wessel2018testing}.  
It is suggested here that some useful intuition about higher cumulants of time series may be built up by interpreting them as representations of the dynamic properties of lower order statistics.
This way of thinking about HOS simplifies the problem of attaching meaning to their representations and may yield some \emph{a-priori} insight into the components recovered through blind identification.

\section*{Acknowledgments}
 Funding from the following sources helped to supported this work:
 
 DOD W81XWH-19-1-0637
 
 NIH R01 NS117753 
 
 NIH 5R01 DC004290-22

\appendix

\section{Higher Order Spectra}
\label{APDX:HOS}
\subsection*{Main properties}
Here we briefly review the main important properties of HOS. These include:
\subsubsection*{Time invariance}
The cancellation of frequencies in \refeqn{EQ:Mts} leads to time-invariance, meaning that arbitrary translations of time will not affect the value of the moment.
The point is trivially apparent in the case of the second-order spectrum, $S(\omega) = \mathrm{E}[|X(\omega)|^2]=\mathrm{E}[X(\omega)X(-\omega)]$, which by discarding phase altogether removes all information related to time-domain structure, yielding back only signal energy as a function of frequency.

\subsubsection*{Phase preservation} 
Spectra of order greater than 2 are not blind to time in the same way but preserve information about temporal structure, up to the discarded time shift. 
In fact, for a transient signal, whose HOS is non-vanishing almost everywhere (by virtue of having a finite duration), it is possible to uniquely recover a waveform from HOS, up to time shift and, in the case of even orders, sign inversion \cite{nikias1993signal,bendory2022signal}.  

\subsubsection*{Indifference to Gaussian noise} 
HOS vanish for Gaussian noise, which follows from the fact that linear, time-invariant Gaussian systems are completely characterized by $1^\text{st}$- and $2^\text{nd}$-order statistics, after which all higher order cumulants, $C^K$, vanish. 
An important consequence of this fact is that consistent estimators of HOS for a non-Gaussian signal are available in the presence of Gaussian noise, which is not case for the power spectrum.   

\subsubsection*{Additivity for additive mixtures}
An especially useful property of cumulant HOS is additivity for additive mixtures of independent processes \cite{cadzow1996}. 
This property makes higher-order cumulants useful for the \emph{blind decomposition} of a non-deterministic HOS into an additive series of deterministic HOS \cite{Kovach2019}, as done here by HOSD.

\subsubsection*{Moment vs. cumulant HOS}
\label{SEC:cumHOS}
The properties, additivity for independent processes and robustness to Gaussian noise, apply to cumulants but not in general to moments. 
Moments are generally easy to compute as they are monomial in form. 
Computing a cumulant is often more cumbersome, as it does not have the simple monomial form described by \refeqn{EQ:Mdef} but involves a summation of products of lower-order moments. 
However, in the frequency domain, this lack of monomial form applies only in those isolated subdomains wherein lower terms of the cumulant expansion degenerate into products of lower-order time-invariant moments \cite{brillinger1967asymptotic,picinbono1999geometrical} (see also Appendix \ref{APNDX:KM} for a simple proof), provided the system is stationary (encompassing conditional non-stationarity as well). These subdomains will be referred to as \emph{degenerate}. For example, in the trispectrum, such a degenerate region exists along $\omega_3=-\omega_1\ne0 \text{ and } \omega_4=-\omega_2\ne0$, for which
\begin{align}
\begin{split}
C^4&(\omega_1,\omega_2,-\omega_1,-\omega_2 ) =\\ 
&\mathrm{E}\left[|X(\omega_1)|^2|X(\omega_2)|^2\right] 
		-  \mathrm{E}\left[|X(\omega_1)|^2\right]\mathrm{E}\left[|X(\omega_2)|^2\right] 
\end{split}
\end{align}
etc. 
For the cumulant HOS of a deterministic process, degenerate subdomains vanish, and therefore must also vanish within the ``true'' cumulant HOS of an ergodic process.
For this reason, discarding degenerate subdomains of moment HOS altogether results in a loss of information that is asymptotically negligible with increasing duration of the analysis window.
Cumulant estimators derived with such a windowing are therefore statistically consistent.

A consequence of this shortcut for computing cumulant HOS in the present application to AM is that the analysis is sensitive to patterns of $2^\text{nd}$-order conditional nonstationarity at a time scale shorter than the analysis window. 
In contrast, previous applications of 4\ts{th}-order statistics towards the identification of conditional nonstationarity have been sensitive to the opposite range of time scales, longer than the analysis window \cite{dwyer1984use, capdevielle1996blind,vrabie2003spectral}.   

\subsubsection*{Deterministic and non-deterministic HOS}
For a deterministic process, whose output is a fixed signal, up to a random time shift, the expectation operator is identity, meaning that its HOS are simply the products of the given signal spectrum at the corresponding frequency multiples and thus factor accordingly.
HOS that factor in this way will be referred to as \emph{deterministic} even if the underlying signal is random with respect to time shifts and non-negative scaling. 
Stochastic systems may also have deterministic HOS. 
In particular, a linear time-invariant (LTI) system driven by stationary white noise has deterministic HOS, a fact which has been exploited for non-minimum phase identification of LTI systems \cite{giannakis1989identification}.

\subsubsection*{HOS inversion}

For systems whose outputs are both \emph{transient} and have \emph{deterministic} HOS, it is not only possible to recover the signal from HOS, but HOS overdetermine the signal, meaning that it is not necessary to estimate HOS over the entire domain to fully recover the signal phase spectrum. 
The minimum number of HOS samples for recovery is on the order of the number of samples in the original signal \cite{bendory2022signal}. 
For example, from the bispectrum, one can reconstruct a transient, deterministic signal from the diagonal axis $X\left(\omega\right)X\left(\omega\right)X^*\left(2\omega\right)$. 
This follows because the support of a transient signal extends across the entire domain of HOS, and recovering the signal from a discretely sampled bispectrum becomes a matter of solving for $N$ unknowns from at least $O(N)$ linearly independent equations.
But this rule does not extend to \emph{non-deterministic} HOS, even when the emitted signals are transient. 
Examples of non-deterministic signals include those that exhibit ``phase-amplitude coupling''' (PAC) between the envelope of a fast oscillation whose envelope, but not phase, correlates with the phase of a lower frequency slow oscillation.
The energy within the bispectrum of such a signal lies within an off-diagonal \cite{Kovach2019}, so cannot be recovered from the the diagonal of the bispectrum.
Likewise the amplitude modulating signals of interest in the present case do not satisfy the deterministic requirement, as the phase of the carrier is assumed to be random with respect to its envelope.

\subsubsection*{HOS of an ergodic process}
As discussed in Appendix \ref{APNDX:mometcum}, any finite observation of an ergodic system can be regarded as a randomly chosen interval from a (typically infinite) deterministic sequence. The assumption of ergodicity thus leads to a view of the ``true'' moments of the system as deterministic, while non-determinism of estimated moments is a consequence of the smoothing and multiplicative distortion in the spectral domain that results from summing over finite observation window. 
This view is conceptually helpful because it allows the statistical properties of an estimator, including bias and error, to be analyzed according to the effects of windowing. 
But it should be emphasized that this view is a conceptual tool for deriving the properties of estimators rather than a statement about the nature of the system under study, which may still exhibit non-deterministic HOS over finite observations windows.

\subsection*{Polycoherence}
Interpretation of HOS is aided by normalizing them to obtain a measure that reflects the statistical magnitude of dependence, often within the range $[0,1]$. 
\begin{align}
\label{EQ:polycoh}
\tilde{M}^K = \frac{\hat{M}^{K}}{\hat{D}^K}
\end{align}
A variety of options for normalization exist \cite{kim1979,shahbazi2014univariate,Kovach2019}, including by the expected standard deviation of the magnitude spectrum for a Gaussian process: 
\begin{align}
\label{EQ:PowerNorm}
 \hat{D}^K_S = \sqrt{\prod_{k=1}^K\hat{S}\left(\omega_k\right)}
 \end{align}
A desirable property of a normalized measure is that its magnitude quantifies the degree of statistical dependence within the range $[0,1]$, from no dependence (0) to perfect dependence (1). 
The normalization given in \refeqn{EQ:PowerNorm} leads to a quantity that may exceed 1. 
Here we will an adopt an alternative normalization according to the expected magnitude:
\begin{align}
\label{EQ:awplvNorm}
 D^K_\mathrm{PLV}=\mathrm{E}\left[|X(\omega_1)||X(\omega_2)|...|X(\omega_K)|\right]
 \end{align}
Plugging in the respective direct estimators for $M$ and $D$ leads to the interpretation of \refeqn{EQ:polycoh} as a weighted phase-locking value between multiple frequencies \cite{hagihira2001practical, Kovach2018,kovach2017biased}.
Alternative normalizations within the unit interval are also possible \cite{kim1979,shahbazi2014univariate}.

\clearpage 

\section{ Moments and cumulants of bounded ergodic processes}
\label{APNDX:mometcum}
\subsection*{Cumulant and moment spectra are equal in non-degenerate subdomains of HOS}
\label{APNDX:KM}
The claim that, for an ergodic and conditionally non-stationary process, moment and cumulant spectra are the same at all frequencies, $M^K(\omega_1,\dots,\omega_K)=C^K(\omega_1,\dots,\omega_K)$, except in ``degenerate'' subdomains, for which $\sum_{j=1}^{J<K} \omega_{i_j} = 0$ for some subset of indices $\{i_j\}$, follows from the general fact that a cumulant of a given order may be written as a polynomial series with the moment of the same order in the leading term and products of two or more lower-order moments and/or cumulants in subsequent terms, whose orders sum to $K$. 
Because the set of $K$ frequency arguments are partitioned among the factors within each term, every term after the first must include a time-varying moment, except when the partitioned subsets of frequencies sum to zero \cite{brillinger1967asymptotic}.

A simple constructive proof may be obtained with an expansion that has at most two factors in each term, a moment and a cumulant, so that each partitioning conveniently involves no more than two subsets of frequencies. 
The expansion is derived from the relation
\begin{align}
\label{EQ:MK}
 M_0e^{-C_0}=1
 \end{align}
where $M_0$ is the moment generating function 
\begin{align}
M_0(S)=\mathrm{E}\left[e^{\int{S(\omega)X(\omega)\mathop{d\omega}}}\right]
\end{align}
 and $C_0$ the cumulant generating function, $C_0(S)=\log M_0(S)$,  where
 \[
 \begin{array}{lll}
 M_K = \frac{d^K}{dS(\omega_1)\dots dS(\omega_K)}M_0 &\text{and}& M^K \equiv M_K(S=0)
 \end{array}
 \]
  and likewise, 
   \[
 \begin{array}{lll}C_K = \frac{d^K}{dS(\omega_1)\dots dS(\omega_K)}C_0&\text{and}&C^K \equiv C_K(S=0)
  \end{array}
 \]
Taking derivatives of both sides of \refeqn{EQ:MK} with respect to $S(\omega_1)$ gives
\begin{align}
\label{EQ:DMK}
M_1(\omega_1) - M_0C_1(\omega_1) = 0 
\end{align}
Further differentiation of \refeqn{EQ:DMK} yields the following relation for the difference between a moment and cumulant of a given order as a series of products between pairs of lower-order moments and cumulants \cite{smith1995}:
\begin{align}
\begin{split}
\label{EQ:MC}
 M^K-C^K =
  \sum_{k=1}^{K-1}\sum_{
 \left\{\{i_p\}, \{j_q\}\right\}_k
 }
  M^k(\omega_{i_1},\dots,\omega_{i_k})C^{K-k}(\omega_{j_1},\dots,\omega_{j_{K-k}})
\end{split}
 \end{align}
The summation  in $\left\{i_p\right\}$  is over  $\binom {K-1}{k}$ unordered partitions of $\left\{2,\dots,K\right\}$ and $\left\{j_q\right\} = \left\{1,\dots,K\right\} \setminus \{i_p\}$. 
For a stationary (including conditionally non-stationary) system, the difference must therefore vanish, unless $\sum_{\{i_p\}_k} \omega_{i_p}=0$ for one or more of the partitions, indicating a degenerate subdomain of frequency space.

\subsection*{Cumulant spectra are proper continuous functions}
\label{APDX:cumulantErgodicity}
The assumption of ergodicity allows estimators derived from an average over repeated serial observations of a single time series to substitute for ensemble averaging over multiple independent realizations of the process, such that the former also converge on the true statistics of the ensemble.
Intervals separated by some sufficiently large lag within a time series may effectively be treated as independent samples from the ensemble. 
This condition can be understood as requiring that the time-domain representation of cumulants to decay to $0$ at large time-lags, such that \cite{brillinger1966computation}
\begin{align}
\int{\left|c^K(\tau_1,\dots,\tau_{K})\right|\left|\tau_j\right|\mathop{d\tau}} < \infty
\end{align}
It follows from ergodicity that cumulant HOS are proper continuous functions \cite{shiryaev1963conditions}. 
Energy associated with any point mass or non-removable discontinuity in the spectral domain is infinitely spread in the time domain, and vice-versa, so that discontinuities in cumulant HOS imply infinitely extended correlations in time, which violates the ergodic assumption.

To show that ergodicity allows consistent estimation of HOS from the observation of a single time series, consider direct window-overlapped segment averaged (WOSA) HOS estimators.
These are obtained by averaging over $N$ intervals from a signal, where each interval is shifted and windowed in a manner that adjusts to the total duration of the observation, $T$, as
\[
x_n (t)= g_T(t)x(t + t_n)
\]
In the spectral domain, windowing amounts a convolution
\begin{align}
X_n(\omega) = \int{G_T(\omega - \eta)X(\eta)e^{i\eta t_n}\mathop{d\eta}}
\end{align}
An estimator for the K\ts{th}-order moment, $\hat{M}^K$, thus relates to the signal moment as:
\begin{align}
\begin{split}
\label{EQ:Mest}
\hat{M}_T^K&(\omega)=
\frac{1}{N} \int\!\cdots\!\int\left(\sum_{n=1}^N e^{i\left(\Sigma_{k=1}^K\eta_k\right) t_n}\right) \prod_{k=1}^K X(\eta_k)G_T(\omega_k - \eta_k)\mathop{d\eta_1}\dots\mathop{d\eta_K}\\
\end{split}
\end{align}
where $\omega = (\omega_1,\dots,\omega_K)$. 
The expectation of \refeqn{EQ:Mest} is given by
\begin{align}
\begin{split}
\label{EQ:MOMest}
\mathrm{E}_X\left[\hat{M}_T^K\right] = 
G^K_T*\left[ M^K \frac{1}{N}\left(\sum_{n=1}^N e^{i\left(\sum_{k=1}^K\eta_k\right) t_n}\right)\right]
\end{split}
\end{align}
where
$
G^K_T(\omega)=\prod_{k=1}^K G_T(\omega_k)
$.
If windows are spaced such that the summation in \refeqn{EQ:MOMest} converges to a unit impulse, or are independently and randomly selected such that the expectation holds
\[
\mathrm{E}\left[\frac{1}{N}\sum_{n=1}^N e^{i\left(\sum_{k=1}^K\eta_k\right) t_n} \right] = \delta(\Sigma_{k=1}^K\eta_k)
\]
then
\begin{align}
\begin{split}
\label{EQ:HOSest}
\mathrm{E}\left[\hat{M}_T^K\right] = G^K_T*\left[\delta(\Sigma_{k=1}^K\eta_k) M^K(\eta) \right]
\end{split}
\end{align}
The expectation of the estimator in \refeqn{EQ:HOSest} converges on $M^K$ within the time-invariant domain provided $G_T\rightarrow \delta(\omega)$.
Procedures for designing estimators that fulfill this requirement are well known \cite{brillinger1967asymptotic}.
The covariance of any two consistent moment estimators vanishes asymptotically, so that
\begin{align}
\begin{split}
\label{HOScov}
\lim_{N\rightarrow \infty} \mathrm{E}\left[\hat{M}_T^A\hat{M}_T^B\right] &= \mathrm{E}\left[\hat{M}_T^A\right]\mathrm{E}\left[\hat{M}_T^B\right]\\
 &= \delta(\Sigma_{k=1}^A\eta_{k})\delta(\Sigma_{k=1}^B\lambda_{k})M^A(\eta)M^B (\lambda) 
\end{split}
\end{align}
But it is also apparent from \refeqn{EQ:Mest} that
\begin{align}
\begin{split}
\label{EQ:HOScov}
\mathrm{E}\left[\hat{M}_T^A\hat{M}_T^B\right] &=  G^{A+B}_T*
\left[ M^{A+B} \frac{1}{N^2} \left(\sum_{n_1=1}^N\sum_{n_2=1}^N e^{i\left(\sum_{k=1}^A\eta_{k}\right) t_{n_1}+i\left(\sum_{k=1}^B\lambda_{k }\right) t_{n_2}}\right)\right]\\
&\rightarrow \delta(\Sigma_{k=1}^A\eta_{k})\delta(\Sigma_{k=1}^B\lambda_{k})M^{A+B}(\eta,\lambda)
\end{split}
\end{align}
representing the degenerate subdomain of moment $M^{A+B}$ in which $\eta$'s and $\lambda$'s vanish separately. 
It must follow that if the process is ergodic and the estimators consistent, then \cite{shiryaev1963conditions} 
\begin{align}
\begin{split}
&M^{A+B}(\eta,\lambda) = 
 C^{A+B}(\eta,\lambda)  + 
\delta(\Sigma_{k=1}^A\eta_{k})\delta(\Sigma_{k=1}^B\lambda_{k})M^A(\eta)M^B(\lambda)
\end{split}
\end{align}
with 
\begin{align}
\label{EQ:cumcondition}
\lim_{\epsilon\rightarrow 0} \int_{\alpha-\epsilon}^{\alpha+\epsilon}{\left|C^{A+B}(\omega)\right|\delta\left(\Sigma_{j=1}^{A+B}\omega_j\right)\mathop{d\omega_k}}= 0
\end{align}
where $\left(\omega_1,\dots,\omega_{k-1},\alpha,,\omega_{k+1},\dots,\omega_{A+B-1}\right)$ lies in a degenerate subdomain.
In fact, because the estimates are asymptotically deterministic, \refeqn{HOScov} must hold for the product of any number of moments.
The same holds in non-degenerate regions: the lower-order moments are not time-invariant and therefore vanish for a stationary process; but $C$ must still be continuous almost everywhere within the plane $\sum_{k=1}^K\omega_k=0$.  
Similarly, estimates of different moment HOS are asymptotically independent. 
Asymptotic independence of the corresponding estimators requires that  the cumulants vanish within all subdomains; were ``vanishing'' is meant in the sense of \refeqn{EQ:cumcondition}, the absence of discrete mass, so that excluding an infinitesimal neighborhood leaves $C^K$ unchanged, except in a region of  infinitesimal energy. For this reason, a moment estimator windowed to exclude degenerate subdomains can be taken as a consistent cumulant estimator:
\begin{align}
 \lim_{T\rightarrow \infty}\hat{C}^K_T=\lim_{N\rightarrow \infty}Q_T\hat{M}_T^K = C^K
\end{align}
provided the support of $Q_T$ converges with the non-degenerate domain of $M^K$.

\clearpage

\section{HOSD Algorithm}
\label{APDX:HOSD_Algorithm}
The HOSD algorithm is motivated by observing that a filter matched to an unknown waveform, $f$,  may be recovered from HOS of a signal that contains the waveform in Gaussian noise at an unknown delay, $x_j(t) = f(t-\Delta t_j) + n_j(t)$,
given the $K^\text{th}$-order deterministic HOS for the waveform
 \[M^K(\omega_1,\dots,\omega_{K-1}) = F(\omega_1)F(\omega_2)\dots F^*(\Sigma_{k=1}^{K-1} \omega_k)\]
using the following relation.
\begin{align}
\label{EQ:PDfilter}
\begin{split}
G_j(\omega_1) = &\int{M^{K*}X_j(\omega_2)\dots X_j^*(\Sigma_{k=1}^{K-1} \omega_k)\mathop{d\omega_2\dots d\omega_K}}\\
= F^*&(\omega_1)\int{F^*(\omega_2)X_j(\omega_2)\dots}F^*(\omega_K)X_j(\omega_K)\\
&\times\delta(\Sigma_{k=1}^K \omega_k)\mathop{d\omega_2\dots\omega_K}\\
=\mathscr{F}&\left\{f \star (f\star x_j)^{K-1}  \right\}
\end{split}
\end{align}
where the $\star$ operator denotes cross-correlation and $\mathscr{F}$ is the Fourier transform.

 The result of the integration in \refeqn{EQ:PDfilter} is the output of a matched filter applied to the signal, $x_j$, raised to the power $K-1$.
 A matched filter produces a peak at the delay of $f$; the exponent applied to the output therefore creates something still more closely resembling an impulse at the delay of $f$, broadening the spectrum of the original output. 
 Because the remaining instance of $F$ outside the integrand in \refeqn{EQ:PDfilter}  is a ``clean'' copy of the target waveform, unaffected by additive noise within $x$, $g_j$ tends to approximate a filter matched to $f$ as it appears in $x$. 
 \begin{align}
\begin{split}
g_j \approx f(\Delta t_j - t)
\end{split}
\end{align}
This relation assumes knowledge of $M^K$.
It is useful in practice because consistent estimators of HOS are available even when $f$ and its delays, $\{\Delta t_j\}$, are initially unknown. 
An estimate of $g$, $\hat{g}$, may therefore be obtained by plugging the estimate $\hat{M}^K$ into \refeqn{EQ:PDfilter}.
Further details of the algorithm and its application to non-deterministic HOS are given in \cite{Kovach2019} and \refapdx{APDX:HOSD_Algorithm}.

The HOSD algorithm is developed from the principle outlined above with the following additional steps. A more detailed explanation of the algorithm and its motivation can be found in \cite{Kovach2019}:
\subsection*{HOS windowing}
The support of the integrand in \refeqn{EQ:PDfilter} does not need to cover the entire HOS frequency domain, but can be restricted or otherwise modified by a non-negative window. 
Such windowing may serve to reduce computational complexity by reducing the number of coefficients needing to be estimated; to emphasize features of interest by restricting estimation to regions of HOS in which the features are most prominent; or to obtain a cumulant estimator through the exclusion of degenerate subdomains in the moment spectrum (see \refapdx{APNDX:mometcum}).
The phase spectrum of the resulting filter estimate will still be matched to that of the target feature within the spectral range encompassed by the window.  
\subsection*{HOS normalization}
In addition to yielding a more direct measure of statistical dependence, normalization of HOS, as used in computing bicoherence or polycoherence generally, effectively optimizes the resulting matched filter according to Fourier domain signal-to-noise ratio.
Different choices of normalization make different assumptions about cross-frequency dependence within the background noise spectrum \cite{Kovach2019}. 
The amplitude spectrum of the ideal matched filter is determined by signal-to-noise ratio \cite{turin1960}; therefore, 
a filter designed with matched phase spectrum, but improperly matched amplitude spectrum is suboptimal with respect to the suppression of noise.
It still fulfills the more essential requirement that the expected output be impulsive  with the correct delay for the target waveform. 
Some related algorithms for estimating channel delays from HOS ignore the problem of matching the amplitude spectrum altogether, \cite{nikias1988time}, relying purely on the phase spectrum, implicitly weighting all frequencies equally.
Although normalization of some type is generally needed, the choice among alternatives does not typically have a dramatic effect on estimates \cite{elgar1988statistics,shahbazi2014univariate}, and in particular should not systematically change the phase spectrum of the output.
\subsection*{Iterative filter estimation}
Iteratively averaging a sample of ``partial delay filters,'' $\{g_j\}$, estimated according to \refeqn{EQ:PDfilter}, applying the resulting average filter to individual records, $\{x_j\}$, realigning the records at the respective maxima of the filter output, efficiently converges on high-quality estimates of the feature waveform and matched filter. 
\subsection*{Power iteration}
Static cumulants are maximized by matched filtering \cite{Kovach2019}, implying a connection between HOSD and blind deconvolution techniques that explicitly maximize static moments such as kurtosis  \cite{wiggins1978,donoho1981minimum,mcdonald2012maximum}. 
Yet the filter estimates obtained from HOSD do not explicitly fulfill this criterion.
For non-deterministic HOS arising from a mixture of processes with deterministic HOS, the delay filter estimates initially contain a mixture of filters matched to distinct components. 
HOSD relies on the iterative realignment on peaks in the filter outputs to selectively converge on one component.
The efficacy of this procedure in separating components associated with similar features depends on noise-related stochastic resonance and other factors, for which reason the result is only an approximation to a moment-maximizing filter.
The initial estimate can however be refined by several rounds of recomputing the delay filter from the feature estimate and re-estimating the feature, which resembles the power iteration technique for recovering the largest eigenvector of a matrix through repeated matrix multiplication.

Power iteration has been applied to the decomposition of $K^\text{th}$-order tensors, typically using a random or arbitrary starting guess \cite{de1995higher}.
But because HOS are restricted to the time-invariant subdomain within the full $K^\text{th}$-order moment tensor, the problem can be regarded as the decomposition of a tensor with a large number of missing entries (namely, all non-time-invariant moments), which degrades the efficacy of standard algorithms \cite{kurucz2007methods}.
While the convergence of power iteration is typically poor when initialized with a random guess in this setting, it succeeds when given a good starting approximation as provided by the previous step.
 
Power iteration applied to HOS can be justified by the following argument: a filter that maximizes the $K^\text{th}$-order moment represents a fixed point in the following Lagrangian function
 \begin{align}
\begin{split}
\Lambda = \int{(g*x)^K(t)\mathop{dt}}  + \lambda\left(1-\int{(g*x)^2(t)\mathop{dt}} \right)
\end{split}
\end{align}
such that
\begin{align}
\label{EQ:LambdaGradient}
\begin{split}
&\mathrm{E}\left[\frac{\partial\Lambda}{\partial |G(\omega_1)|}\right] =\\
K\int\dots\int& M^KG(\omega_2)\dots G(\omega_K) 
\delta\left(\Sigma_{k=1}^K\omega_k\right)\mathop{d\omega_2\dots d\omega_K} \\
 -& 2\lambda G^*(\omega_1)|X(\omega_1)|^2 =0
\end{split}
\end{align}
Allowing $G$ to be whitened by the spectrum of X,
\[G_X(\omega) =\left|X(\omega)\right|G(\omega)\]
under small perturbations around $G$, the following iteration may be expected to converge to the fixed point: 
\begin{align}
\label{EQ:PICA}
\begin{split}
\hat{G}_X^{*(n+1)}&(\omega_1) =\\
\int\dots\int& \tilde{M}^K\hat{G}_X^{(n)}(\omega_2)\dots \hat{G}_X^{(n)}(\omega_K) 
\delta\left(\Sigma_{k=1}^K\omega_k\right)\mathop{d\omega_2\dots d\omega_K}
\end{split}
\end{align}
where $\tilde{M}$ is normalized according to some estimate of the product of amplitude spectra of $X$, and $\hat{G}_X$ rescaled at each iteration to explicitly satisfy the constraint of unit energy in the output (hence the constant terms in \refeqn{EQ:LambdaGradient} may be ignored). 
\subsection*{Thresholding}
The relationship of HOS to static cumulants of corresponding order leads to a heuristic criterion for thresholding the filter output: the threshold is set such that the residual static $K^\text{th}$-order cumulant vanishes, or otherwise attains a desired tradeoff between sensitivity and specificity according to the expected null distribution of the cumulant for Gaussian noise. 
\subsection*{Component reconstruction}
For filters estimated from normalized HOS estimates, the result of thresholding directly approximates a weighted probability distribution for $\Delta t$ \cite{Kovach2019}.
A reconstructed component signal may therefore be obtained by convolving the estimated feature waveform with the thresholded filter output and scaling the result to minimize squared error.  

\subsection*{Decomposition through serial deflation}
Subtracting the reconstructed component from the original signal and repeating the procedure on the residual yields an additive decomposition of HOS (HOSD). 
Deflation may be repeated until a stopping criterion is met, such as recovery of a predetermined number of components or attainment of a threshold in the static $K^\text{th}$-order cumulant of the detection filter output.

\subsection*{Implementation}
\label{SEC:HOS_METHODS}
\subsubsection*{HOS estimation}
Applying the algorithm summarized above to discretely sampled data involves the standard considerations and caveats when moving from continuous to discrete time.
After  line noise removal and artifact rejection, data were segmented according to (\ref{EQ:DFT}).
Segments were tapered with Sasaki's window function \cite{sasaki1975minimum}.
HOS computed with the DFT are invariant to circular shifts of the weighted signal within the analysis windows, but not in general to linear shifts in the original signal.

The direct approach to HOS estimation follows the same procedure as the WOSA technique for power spectral estimation \cite{welch1967use}. 
HOS coefficients are  estimated by averaging the corresponding product of Fourier coefficients across segments:
\begin{align}
\begin{split}
\label{EQ:directHOS}
\hat{M}^K[m_1,\dots,m_{K-1}] &= \frac{1}{J}\sum_{j=1}^J\prod_{k=1}^K X_j[m_k]
\end{split}
\end{align}
where $m_K = \left(-\sum_{k=1}^{K-1}m_k\right) \mod P$.
In the present application, $K=4$.

As with the power spectrum, WOSA estimates are consistent \cite{brillinger1967asymptotic} meaning that $P$ and $J$ can both be increased as a function of total signal length, $N$, to obtain an asymptotically converging and unbiased estimate, provided the system under study is bounded and ergodic. 
Note however that HOS estimates computed from the DFT in this way are not in general shift invariant under linear or circular time shifts of the original time series.
This lack of shift invariance reflects the fact that the selection of windows necessarily entails some localization in time.
Specifically, if $\Delta_j = j\Delta P\mod N$ for some constant interval, $\Delta P$, they will be invariant under shifts modulo $\Delta P$.
Estimators obtained by averaging over segment HOS can be made shift invariant over the whole time series by densely spacing the analysis windows, setting $\Delta P=1$.
Nevertheless, deviations from strict shift invariance for $\Delta P > 1$, as when $\Delta P = \frac{1}{2}P$, are likely to be a minor concern compared to the computational expense and extreme redundancy of maximally dense window spacing, particularly when the segmentation can be assumed random with respect to the signal. 
This point applies also to ordinary WOSA power spectral estimates and is not unique to HOS estimation.
Windowing may be used deliberately to preserve timing information in the form of  time-varying or event-related HOS estimates, as is commonly done for spectral power in computing a spectrogram.

\subsubsection*{Normalization and HOS Windowing}
HOS estimates were normalized to yield polycoherence, which more directly represents the statistical relationship between frequencies, while also optimizing the  signal-to-noise ratio of subsequent filter estimates. 
The present applications have relied on mean magnitude normalization \cite{hagihira2001practical,kovach2017biased}:
\begin{align}
\begin{split}
\hat{D}^K[m_1,\dots,m_{K-1}] = \frac{1}{J}\sum_{j=1}^J\prod_{k=1}^K \left|X_j[m_k]\right| 
\end{split}
\end{align}
although several alternative normalizations might be equally justified \cite{shahbazi2014univariate}.

Additional modification of HOS weighting through the application of a non-negative window function, $Q$ in the HOS domain is also often desirable.
Reasons include obtaining cumulant HOS by suppressing degenerate slices within the moment spectrum, reducing computational overhead by limiting the number of coefficients needing to be estimated, and restricting estimators to regions within HOS relevant for some particular question. 
\begin{align}
\label{EQ:discrete_polycoh}
\begin{split}
\tilde{M}^K= \frac{\hat{M}^K}{\hat{D}^K}Q\left[m_1,\dots,m_K\right]
\end{split}
\end{align}
In the present application, Q windows within the diagonal slice:
\begin{align}
Q[m_1,m_2,m_3] = 
\left\{
	\begin{array}{ll}
		1 & \text{if } m_i=m_j\ne-m_k \ne0 \\
		0 & \text{otherwise}
	\end{array}
	\right.
\end{align}
where $i\ne j\ne k$.
The support of Q was further modified to include only a specified range of carrier and modulation frequencies.
In practice, only the subset of coefficients within the support of Q need to be estimated, avoiding the computational expense of full trispectrum estimation.

\subsubsection*{Filter estimation}
For each segment of data, a partial delay filter estimate was obtained following \refeqn{EQ:PDfilter} as
\begin{align}
\begin{split}
G_j[m_1]= \sum_{m_2=1}^P\cdots \sum_{m_{K-1}=1}^P  \frac{\tilde{M}^{K*}[m_1,\dots,m_{K-1}]}{\hat{D}^K[m_1,\dots,m_{K-1}]}\prod_{k=2}^KX[m_k]
\end{split}
\end{align}
The HOSD algorithm recovers a filter through iterative circular alignment and averaging of the segment partial delay filters
 \begin{align}
\begin{split}
\hat{G}^{(n+1)}[m]= \sum_{j=1}^Js_jG_j[m]e^{-2\pi i\frac{m}{P} \Delta p^{(n)}_j}
\end{split}
\end{align}
where 
\[ \Delta p^{(n)}_j = \left\{
\begin{array}{ll}
\argmax_{p} \;\hat{g}^{(n)}*x_j &\text{ for } K \text{ odd} \\
\argmax_{p} \;\left|\hat{g}^{(n)}*x_j\right| & \text{ for } K \text{ even}
\end{array}
\right.
\]
and
\[ s_j = \left\{
\begin{array}{cl}
1&\text{ for } K \text{ odd} \\
\mathop{\mathrm{sgn}}\;\left(\hat{g}^{(n)}*x_j\right)\left[\Delta p^{(n)}_j\right] & \text{ for } K \text{ even} 
\end{array}
\right.
\]

\subsubsection*{Feature estimation}
An estimate of the signal feature to which $\hat{G}$ is matched was obtained directly by averaging the realigned segments:
 \begin{align}
 \label{EQ:shiftAv}
\begin{split}
\hat{F}[m]= \sum_{j=1}^Js_jX_j[m]e^{-2\pi i \frac{m}{P} \Delta p_j}
\end{split}
\end{align}

\subsubsection*{Power iteration}
Once stopping criteria for the previous step were attained (convergence within a maximum of 25 iterations), the estimate of $\hat{G}$ was further refined according to \refeqn{EQ:PICA} through the following iteration:
 \begin{align}
 \label{EQ:PowerIter}
\begin{split}
\hat{G}^{(n+1)}[m_1]= \sum_{m_2=1}^P\cdots \sum_{m_{K-1}=1}^P  \frac{\tilde{M}^{K^*}}{\hat{D}^K}\prod_{k=2}^KF^{(n)}[m_k]
\end{split}
\end{align}
where $\hat{F}^{(n)}$ is updated as the shifted average in \refeqn{EQ:shiftAv}.
The iteration was repeated for a fixed number of 50 iterations.

\subsubsection*{Filtering}
The filter resulting from the previous step was applied to the original input time series, $u = \hat{g}*x$. 

\subsubsection*{Thresholding}

As explained in \cite{Kovach2019}, 
the thresholded output of the component filter can be treated as an approximation of the probability distribution of feature occurrence times, weighted by amplitude, in additive noise. 
A heuristic argument justifies setting the threshold, $\hat{u} = u\cdot I(u^K>\theta)$, such that the static cumulant of the residual vanishes,
$ \theta : \mathrm{cum}_K\left\{u(u^K\le\theta)\right\} = 0$.

\subsubsection*{Signal reconstruction}
Because the thresholded filter output, $\hat{u}$, approximates a weighted probability distribution of feature occurrence times, a reconstructed signal, approximating an expectation, was obtained by convolving the result of the previous stop with the feature waveform and scaling to minimize squared error. 
 \begin{align}
 \label{EQ:Xrec}
\begin{split}
\hat{x}= b\hat{f}*\hat{u}
\end{split}
\end{align}
where $b = \underset{b}{\argmin} \lVert x-b\hat{f}*\hat{u}\rVert$
\subsubsection*{Deflation}
Subsequent components were estimated by repeating HOSD on the residual, $x_\mathrm{res.} = x-\hat{x}$. HOSD was repeated until the value of the static cumulant (kurtosis) for the filtered output, $u$, fell under 0.1 for two successive components.

\begin{figure*} 
\includegraphics{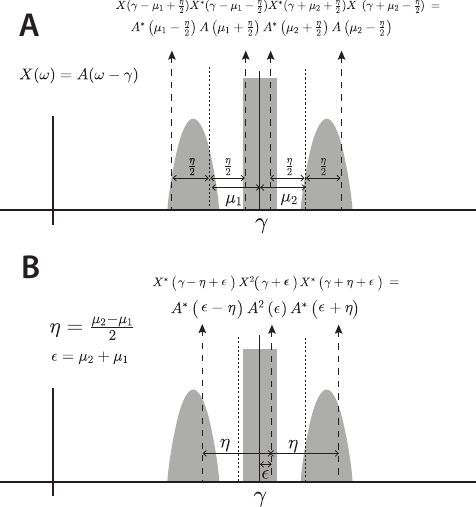}
\caption{\label{fig:band_illo} 
{\bf A}: Illustration of the trispectrum of an amplitude modulated signal as the cross spectrum between adjacent bands of the modulation spectrum. {\bf B}: The two dimensional ``diagonal slice'' subdomain conveys the interaction between upper and lower sidebands and the carrier band.
 }
\end{figure*}

\begin{figure*} 
\includegraphics[width=\textwidth]{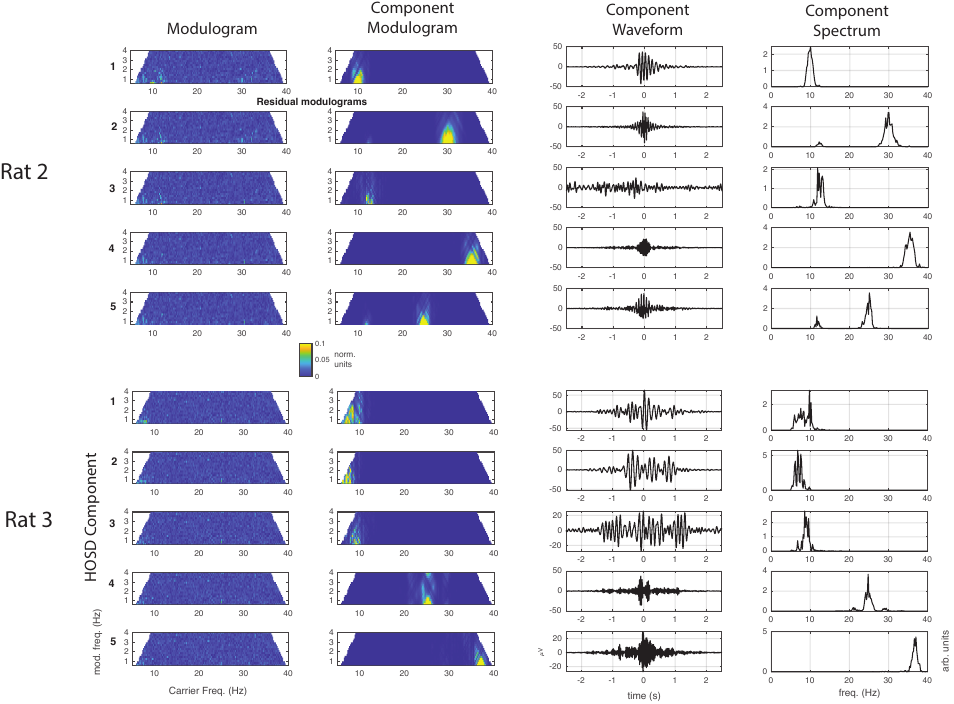}
\caption{\label{fig:HOSD_rat2_and_3} 
Performance with weak oscillatory bursting. Modulograms provided weak evidence for oscillatory bursting in data from rat 2 and no clear indication of bursting in rat 3. HOSD nevertheless recovered oscillatory features in both cases, suggesting the presence of weak oscillatory bursting near the limit of detectability.
}
\end{figure*}

\bibliography{Bibliography}

\end{document}

%% file: abstract.tex
A need to identify modulation of spectral power arises frequently in the analysis of physiological signals. Estimation of power in the relevant bands through filtering and envelope extraction has several limitations: the choice of filter may bias any resulting estimate, while additive Gaussian noise becomes non-Gaussian due to the nonlinearity of envelope computation. The present work considers how spectral decompositions of higher-order cumulants (higher-order spectra, HOS) avoid these limitations, with an emphasis on the use of the trispectrum to identify modulated oscillations. Specifically, it is shown: 1) The trispectrum may be interpreted as a measure of linear dependencies of power across frequencies by viewing it as the cross spectrum of the Wigner-Ville distribution. 2) A particular two-dimensional subdomain of the trispectrum is useful for identifying modulated carriers, recovering essential spectral properties of both the modulating and carrier signals while avoiding the cubic complexity of full trispectrum estimation. A representation of this subdomain, the modulogram, is demonstrated as a tool for identifying and distinguishing different forms of modulation. 3) As a cumulant-derived measure, the modulogram is not biased by additive Gaussian noise. 4) Modulogram phase retains information by which temporal patterns of modulation may be identified. 5) A recently described additive decomposition of HOS (HOSD) further aids identification when applied to the trispectrum. These developments are illustrated with the blind detection of beta bursts in rodent and human local field potential recordings. Finally, the relationship between the present approach and prior techniques of blind identification (BI) through moment maximization, including blind deconvolution and independent component analysis, is considered.